\newcommand{\Om}{\Omega}			% Shift in frequenza
\newcommand{\q}{\vec{q}}                              %q-vector
\newcommand{\rr}{\vec{r}}                               %r-vector
\newcommand{\ex}{\vec{e}_x}
\newcommand{\ey}{\vec{e}_y}
\newcommand{\ez}{\vec{e}_z}
\newcommand{\DD}{{\mathcal D}}                           % Phase mismatch general
\newcommand{\D}{{D}}                                         % Phase mismatch  plane wave pump 
\newcommand{\Gone} { \vec{G}_1}
\newcommand{\Gtwo} { \vec{G}_2}

\newcommand{\Gx} { \vec{ \mathcal G}_x}  %vettore Gx in 3D

\newcommand{\A} {\hat{A}}
\newcommand{\As} {\hat{A}_s}
\newcommand{\Ai} {\hat{A}_i}

\newcommand{\ah} {\hat{a}}
\newcommand{\as} {\hat{a}_s}

\newcommand{\bs} {\hat{b}_s}
\newcommand{\bi} {\hat{b}_i}
\newcommand{\cs} {\hat{c}_s}
\newcommand{\ci} {\hat{c}_i}

\newcommand{\sigmas} {\hat{\sigma}_s}
\newcommand{\sigmai} {\hat{\sigma}_i}
\newcommand{\deltas} {\hat{\delta}_s}

\newcommand{\apiu} {\hat{a}_{i+}}
\newcommand{\ameno} {\hat{a}_{i-}}
\newcommand{\azero} {\hat{a}_{s0}}
\newcommand{\auno} {\hat{a}_{i1}}
\newcommand{\adue} {\hat{a}_{i2}}
\newcommand{\apiuout} {\hat{a}_{i+}^{\mathrm out} }
\newcommand{\amenout} {\hat{a}_{i-}^{\mathrm out}}
\newcommand{\asout} {\hat{a}_{s0}^{\mathrm out}}
\newcommand{\apiuin} {\hat{a}_{i+}^{\mathrm in} }
\newcommand{\amenin} {\hat{a}_{i-}^{\mathrm in}}
\newcommand{\asin} {\hat{a}_{s0}^{\mathrm in}}

\newcommand{\aunoout} {\hat{a}_{i1}^{\mathrm out}}

\newcommand{\adueout} {\hat{a}_{i2}^{out}}

\newcommand{\Ug}{U_\gamma}
\newcommand{\Vg}{V_\gamma}

\newcommand{\w}{\vec{w}}			% 3-D q,omega

\newcommand{\vxi}{\vec{\xi}}	
	
\newcommand{\sinc}{{\rm sinc}}

\newcommand{\nn}{\nonumber}
\newcommand{\bsub}{\begin{subequations}}
\newcommand{\esub}{\end{subequations}}
\newcommand{\beq}{\begin{equation}}
\newcommand{\eeq}{\end{equation}}
\newcommand{\beqa}{\begin{eqnarray}}
\newcommand{\eeqa}{\end{eqnarray}}
\newcommand{\beql}{\begin{subequations}\begin{eqnarray}}
\newcommand{\eeql}{\end{eqnarray}\end{subequations}}
\documentclass[pra,aps,amsmath, nofootinbib,onecolumn,notitlepage,showpacs,longbibliography]{revtex4-1}
\usepackage{amsmath} 
\usepackage{empheq}
\usepackage{amssymb}
\usepackage{graphicx}
\usepackage{longtable}
\usepackage{float}   % per posizionare meglio le figure con [H]
\usepackage{verbatim}  %multi-line comment \begin{comment}....\end{comment}
\usepackage{braket}
\usepackage{color}
\begin{document}
\title{Golden Ratio entanglement in hexagonally poled nonlinear crystals}
\author{ Alessandra~Gatti$^{1,2}$ , Enrico~Brambilla$^2$, Katia Gallo$^3$ and Ottavia Jedrkiewicz$^{1,2}$ }
\affiliation{$^1$ Istituto di Fotonica e Nanotecnologie of CNR, Piazza Leonardo Da Vinci 32, Milano, Italy;  
$^2$ Dipartimento di Scienza e Alta Tecnologia, Universit\`a dell'Insubria, Via Valleggio 11,  Como, Italy,
$^3$ KTH - Royal Institute of Technology, Roslagstullsbacken 21, 10691 Stockholm, Sweden}
\email{Alessandra.Gatti@mi.infn.it}
\begin{abstract}
This work analyses   the quantum state of twin photons and twin beams generated 
by parametric-down conversion 
in a hexagonally poled  photonic  crystal, characterized by the simultaneous presence of two 
nonlinear processes sustained by
 two  vectors of the reciprocal lattice. 
%encompassing both the photonic entanglement generated for low gains and  the continuos-variable entanglement.  
 In those special points of the fluorescence spectrum   where the two processes coexist,  we show that a tripartite entangled state is realized, equivalent to a single parametric process followed by a beam-splitter. 
 By proper angle tuning a peculiar resonance condition is reached, with a transition to a  4-mode entanglement, dominated by  the {\em Golden Ratio} of the segment $\phi= (1+\sqrt 5)/2$. A maximal coherence between the two nonlinear processes is here estabilished, as the overall process is shown to be equivalent to {\em two} independent parametric processes followed by a beam-splitter.  We offer an interpretation of the occurrence of the golden ratio in this  system  based on an analogy between the  evolution of the light modes and the Fibonacci sequence.
 %is the famous Golden Ratio equivalent to i) two independent parametric processes, with unbalanced gains $g_0\phi$ and $-g_0/\phi$, where $\phi= (1+\sqrt 5)/2$ is the famous Golden Ratio, ii) followed by a  beam-splitter that mixes the two processes according to the Golden Ratio. We  offer an interpretation of the occurrence of such particular number,  based on the microscopic processes taking place.
\end{abstract}
%\centerline{Version \today}
\maketitle
%%%%%%%%%%%%%%%%%%%%%%%%%%%%%%%%%%555
\section*{Introduction}
\label{sec:intro}
Nonlinear photonic  crystals, characterized by a two-dimensional periodic modulation of the nonlinear response \cite{Berger1998,Arie2007,Arie2009}, 
offer a high degree of flexibility for engineering the  properties of  optical parametric processes, because of the multeplicity of   vectors of the nonlinear lattice providing quasi phase matching.  When considering the generation of twin photons or twin-beams by parametric  down-conversion  (PDC), these photonic crystals have shown interesting potentialities as monolithic sources  of  path entangled photonic states \cite{Gong2012, Jin2013, Megidish2013},  and  may provide novel compact schemes for continuous-variable quantum technologies \cite{Gong2016}. 

 In this work we  analyse  the quantum state of twin photons and twin beams generated in a hexagonally poled nonlinear photonic  crystal (HexNPC) with a quadratic nonlinearity (see \cite{Gallo2011, Levenius2012, Jin2013}). 
%On the one side,  we offer  an extensive and novel view of the tripartite entangled state  thereby produced, on the other we describe a peculiar 4-mode entangled state generated by proper angle tuning of the input pump. 
The HexNPC is  characterized by the coexistence of two nonlinear processes, sustained by the two fundamental vectors of the  reciprocal  lattice of the nonlinearity (Fig.1). 
In the spectral-angular domain of the down-converted  light there are special points where phase matching occurs simultaneously for both processes, and where  twin photons  may originate by either processes. In the high-gain regime  the two possibilities add coherently and stimulate each other,  giving rise to unusual isolated hot spots in the parametric emission \cite{Liu2008,Chen2014,Conforti2014}.
\par
In the quantum domain, we describe  a general  scenario of tripartite entanglement holding among specific triplets of hot-spots.  We show that here
%, corresponding in the high-gain to as many hot-spots. 
the action of the  photonic crystal is  equivalent to a single nonlinear  process generating a pair of entangled twin beams, followed by a 50:50  beam-splitter  dividing one of the twin beams into two separated paths.   The  occurrence of such a situation was
already suggested in  \cite{Jin2013}. In contrast to  the analysis  performed in \cite{Jin2013},  which was limited to the two-photon state generated in the spontaneous regime, our study is valid for any photon number,  encompassing both  the photonic and continuous-variable entanglement,  and emphasizes the role of conditional measurements on generating path entanglement. 
%%%%%
\begin{figure}[thb]
\includegraphics[scale=0.5]{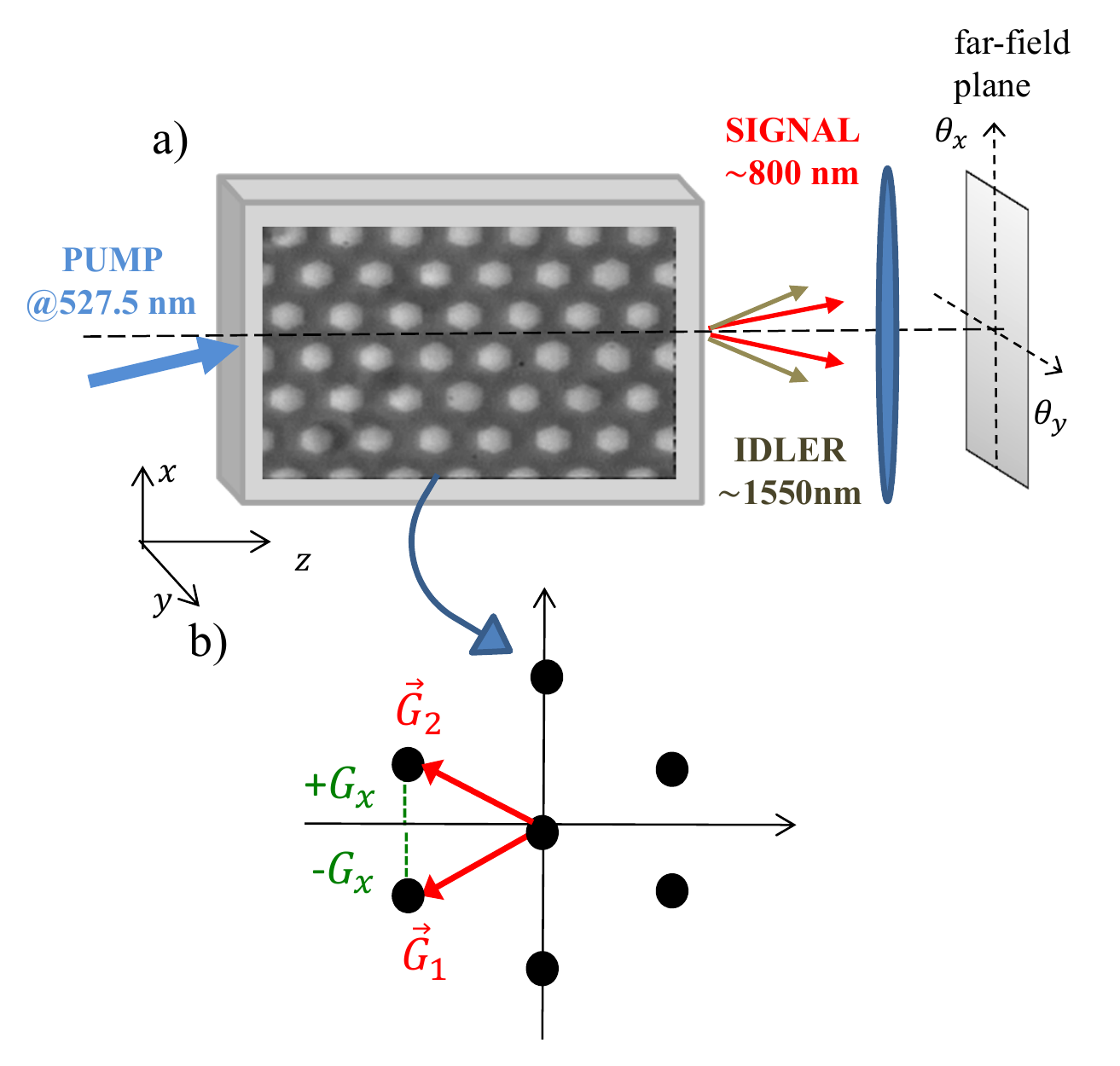}
\caption{(a) Scheme of parametric down-conversion in a hexagonally poled nonlinear photonic crystal. The pump beam is allowed to be slightly tilted  with respect to the  symmetry axis $z$ of the nonlinear pattern.
 (b)  Vectors of the reciprocal lattice contributing to quasi phase-matching.
%c) Numerical simulations of the far-field distributions of the signal and idler fields when  the pump is tilted to  super-resonance, in the nondegenerate configuration of \cite{Gallo2018}.
}
\label{fig1}
\end{figure}
%%%%%%%%%%%%%%%%
\par
In a  related experimental work \cite{Gallo2011}, we observed that by properly tuning the angle of incidence of the pump laser, it was possible to reach a particular condition, that was referred to as  {\em superresonance}.  In such condition, 
two triplets of hot spots coalesce into four coupled modes,  with a sudden enhancement of  the brightness of the hot-spots. Indeed,  we demonstrated  that the rate of  growth  of parametric light along the crystal increases in these modes by   the  famous { \em Golden Ratio} of the segment, $ \phi= \frac{1 + \sqrt{5}}{2}$.  In the present  work we show that this condition, corresponding to a transverse  spatial resonance between the pump and the nonlinear lattice,  gives rise to a quadripartite entangled state, and enables  the maximal coherence between the two concurrent nonlinear processes characterizing the hexagonal photonic crystal. In this condition we demonstrate that the action of the HexNPC  is equivalent to i) {\em two}  independent parametric  processes, with different gains $ g_0 \phi$  and $ \frac{-g_0}{\phi}$, generating two pairs of entangled twin beams, ii) followed by an unbalanced beam-splitter that mixes the two processes according to the Golden Ratio. 
\par An original aspect of our analysis is that it fully takes into account the 3D character of the parametric emission (transverse spatial coordinates plus temporal frequency), by means of extended 3D+1 numerical simulations of the device, complemented by approximated analytical models.  This allows us to show that both the 3-mode and 4-mode entanglement can be realized at different frequencies over a broad bandwidth,  opening the possibility of a widely tunable implementation  of novel quantum states of light. 
%We offer a physical picture of the occurrence of such particular number, based on the microscopic processes taking place.
\section{The model}
\label{sec:model}
We consider the geometry depicted in  Fig.\ref{fig1}, where a laser pump beam propagates in a   $\chi^{(2)}$ photonic crystal, with a hexagonal pattern of the nonlinear response \cite{Gallo2011, Levenius2012, Jin2013,Stensson2014,Conforti2014}, generating signal and idler waves at lower energies.  Light is assumed to propagate mainly 
%be confused with the Z crystalline axis)
%pointing along the y-axis of our frame of reference) 
along the symmetry axis of the  pattern ($z$-axis of our frame of reference), but we allow the input pump to be slightly tilted in the pattern plane ($(x,z)$ plane). Although our analysis  may apply to different tuning conditions,
we  focus
% on a hexagonal pattern of the nonlinearity  and 
on the  type 0 process, where all the three waves have the same extraordinary polarization, and on non-degenerate parametric emission, in a configuration similar to the experiment in  \cite{Gallo2018}.
%and then specialize the results to the degenerate configuration,  
%simlar to that in Ref.\cite{Jin2013} more suitable for generating quantum entanglement and squeezing.
\par
 The model is formulated in terms of coupled propagation equations for the pump  signal and  idler  field operators, describing three wave-packets centered around frequencies $\omega_p$, $\omega_s$ and $\omega_i= \omega_p-\omega_s$,respectively. 
%for which QPM occurs for some  directions of propagation that will be specified later on. 
The grating of the nonlinearity  is described by keeping only the leading order terms in the Fourier expansion of the nonlinear-susceptibility \cite{Berger1998,Levenius2013}
\beq
d(x,z) \simeq e^{-iG_z z}\left[ d_{01} e^{-i G_x x} + d_{10} e^{i G_x x} \right] = 2 d_{01}  e^{-iG_z z} \cos{(G_x x)}  ,
\eeq
where only the contribution of the two fundamental vectors of the reciprocal lattice (Fig.\ref{fig1}b), 
$\Gone \equiv \vec{G}_{01}= -G_x \ex -G_z \ez$ and $\Gtwo \equiv \vec{G}_{10}= +G_x \ex -G_z \ez$,   which  allow  quasi-phase matching (QPM), have been retained. For the  hexagonally poled Mg-doped Lithium Tantalate (LiTaO$_3$) crystal used in \cite{Gallo2018,Jin2013},  $d_{10}=d_{01} \simeq 0.29 d_{33}$ ($d_{33}=17\,$pV/m), $G_z=\frac{2\pi}{\Lambda}$ and $G_x=\frac{2\pi}{\sqrt{3}\Lambda}$ where  $\Lambda%=8.3\,\mu$m
$ is the poling period.
\par
The propagation equations are best written in the Fourier domain spanned by the 3D vector $\w=(\q,\Omega)$, where $\q=q_x \ex + q_y \ey $ is  the wave-vector in the plane transverse to the mean direction of propagation $z$, and $\Om$ is the frequency shift with respect to the carrier frequencies.  The conjugate spatio-temporal domain is in turn described  by the vector $\vxi= (\rr,t)$,  $\rr=x \ex + y \ey$, with the convention for the inner product $ \w\cdot \vxi  := \q \cdot \rr -\Om t$ .  We consider the three slowly varying  envelope operators 
$\A_j (\w,z)  \propto e^{-i k_{jz} (\w) z} \int \frac{d^3 \xi} {(2\pi)^{\frac{3}{2} }}  e^{-i \xi \cdot \w}  \hat E_j^{(+)} (\xi,z) e^{i \omega_j t} 
$ for the signal (j=s), the idler (j=i) and the pump (j=p) fields,  where  $ \hat E_j^{ (+)}$ are the positive-frequency parts of the respective electric field operators, and 
$k_{jz} (\q,\Om) = \sqrt{k_j ^2(\q, \Om)  -q^2}$ are the z-components of the wave-vectors, with the wave number
 $k_j (\q,\Om) = n_j(\q, \Om) \frac{\omega_j +\Om}{c} $ being determined by the linear dispersion relation of the j-th wave in the medium. Dimensions are such that $\hat A_j^\dagger (\w) \hat A_j (w)$ are photon  numbers per unit frequency and wavevector squared. 
The field operators $\A_j$ are slowly varying along $z$, because all the effects of the linear part of the interaction with the medium, contained in $k_{jz}$,  have been subtracted.
% because they evolve only under  the effect  the $\chi^{(2)}$ interaction \cite{Gatti2003}, 
Their evolution along the nonlinear photonic crystal  is described by the following equations  (see  \cite{Gatti2003,Brambilla2014} for  derivations of similar equations): 
\begin{align}
\frac{\partial}{\partial z}   \hat{A}_s  (\w_s, z )   &=  \chi \int 
 \frac{d^3 \w_p }  {(2\pi)^{\frac{3}{2}} }  \hat{A}_p(\w_p,z) 
\left[   \hat{A}_i^\dagger(\w_p -\w_s -\Gx, z)  e^{-i \DD_1 (\w_s, \w_p ) z }  +
%\right. \nn \\ + & \left.   
\hat{A}_i^\dagger(\w_p -\w_s +\Gx, z)  e^{-i \DD_2 (\w_s, \w_p ) z }  \right]
%&\left[   \hat{b}^\dagger(\w_p -\w_s -\Gx, z)  e^{-i \DD_1 (\w_s, \w_p -\w_s-\Gx) z } \right. \nn \\ + & \left.   \hat{b}^\dagger(\w_p -\w_s +\Gx, z)  e^{-i \DD_2 (\w_s, \w_p -\w_s +\Gx) z }  \right]
\label{NLs}\\
\frac{\partial}{\partial z}   \hat{A}_i (\w_i, z )   &=  \chi \int 
 \frac{d^3 \w_s }  {(2\pi)^{\frac{3}{2}} }   \hat{A}_p (\w_p,z) 
\left[  \hat{A}_s ^\dagger(\w_p -\w_i -\Gx, z)  e^{-i \DD_1 (\w_p -\w_i -\Gx, \w_p) z }  + \right. \nn \\
  &  \qquad \qquad \qquad \qquad \qquad \qquad \qquad + \left.  \hat{A}_s^\dagger(\w_p -\w_i +\Gx, z)  e^{-i \DD_2 (\w_p -\w_i +\Gx, \w_p) z }
 \right]
\label{NLi} \\
\frac{\partial}{\partial z}    \hat{A}_p  (\w_p, z )   &= - \chi \int 
 \frac{d^3 \w_s }  {(2\pi)^{\frac{3}{2}} }   \hat{A}_s (\w_s,z) 
   \left[ \hat{A}_i  (  \w_p  -\w_s-\Gx, z)  e^{i \DD_1 (\w_s, \w_p ) z } +
% \right. \nn \\ + & \left.  
 \hat{A}_i  (  \w_p -\w_s +\Gx, z)  e^{i \DD_2 (\w_s, \w_p ) z }  \right]
\label{NLp}
\end{align}
where $\Gx= (G_x, 0,0)$  is a short-hand notation for the x-component of the reciprocal lattice vector in the 3D Fourier  space, and
$
\chi
%\sqrt{\frac{ \hbar (\omega_s +\Om_s)^2 ( \omega_i +\Om_p +\Om_s)^2 ( \omega_p + \Om_p)^2   }{ 8\epsilon_0 c^6 
%k_{sz} (\Om_s,\q_s) k_{pz} (\Om_p,\q_p ) k_{iz} ( \q_p -\q_s \pm\Gx,  \Om_p -\Om_s  }}  
\simeq d_{01}
\sqrt{\frac{ \hbar \omega_s   \omega_i  \omega_p   }{ 8\epsilon_0 c^3 n_i n_s n_p  }}  
$. 
The first and second term at r.h.s of these equations  describe all the possible three-photon interactions $\w_p \longleftrightarrow \w_s \, \w_i$ 
that satisfy the generalized  energy-momentum conservation by means of  the  lattice vectors $\Gone$ and $\Gtwo$, respectively : 
\bsub
\label{phase-matching}
\begin{align}
&\Om_s + \Om_i  = \Om_p  \quad &\text{energy conservation} \\
&\q_s + \q_i =\q_p \mp G_x \ex  \quad &\text{transverse momentum conservation} \\
 &k_{sz} +  k_{iz}  =  k_{pz} - G_z  \quad &\text{longitudinal momentum conservation}
\end{align}
\esub
The last rule, rigorously obeyed only for an infinite propagation length,  is accounted for by  the QPM functions: 
\begin{align}
 \DD_{1}(\w_s,\w_p)=k_{sz}(\w_s)+k_{iz}(\w_p - \w_s  -\Gx)-k_{pz}(\w_p)+G_z  \nn\\
\DD_{2}(\w_s,\w_p)=k_{sz}(\w_s)+k_{iz}(\w_p - \w_s  + \Gx)-k_{pz}(\w_p)+G_z 
\label{DD12}
\end{align}
which contain the effects of temporal dispersion and diffraction at any order.  
% In the degenerate case  the signal and idler fields have the same carrier frequency and polarization and  cannot be distinguished. Thus,  this configuration is described by Eqs. \eqref{NLs} and \eqref{NLp} with the substitution $\Ai \to \As$. 
\par
These equations are in general too complicated to be solved without approximations, but stochastic simulations can be performed in the medium-high gain regime of PDC in the framework of the Wigner representation, where the field operators are replaced by c-number fields \cite{Gatti1997}. In this context, the vacuum fluctuations at  the crystal entrance facet  are  simulated by Gaussian white noise. The propagation equations are then  integrated with a pseudo-spectral method, splitting linear propagation, solved in Fourier space, from nonlinear propagation solved in direct space.  The linear part of the evolution is evaluated by using empirical Sellmeier formulas, found in \cite{Lim2013} for the LiTaO$_3$ crystal.  Quantum expectation values (in symmetric ordering) can be obtained by averaging over the initial conditions. A single stochastic realization can be taken as a semiclassic simulation of the system. Indeed, 
despite the unavoidable limitations of the size of the numerical grid (typically 512 x 256 x512 points in the x, y and t directions), such simulations were able to closely   reproduce,  also quantitatively,  the classical features of optical parametric generation  in a HexNPC observed in \cite{Gallo2018}.
%%%%%%%%%%%%%%%
\section{Parametric limit and shared modes}
Analytic results can be derived in the {\em parametric limit} where the pump beam, undepleted by the down-conversion process,  is approximated as a classical plane-wave of constant amplitude  along the sample. 
In this limit, by  assuming that the input  pump  propagates  in the $(x,z)$ plane at an angle $\theta_p \simeq \frac{q_p}{k_p} \ll 1 $ with the $z-$axis,  we can set  $\hat A_p (x,y,t,z) \to \alpha_p e^{i q_p x} $ where $\alpha_p$ is the classical  field amplitude. Linear  evolution equations for the signal and idler operators 
are  then obtained from  Eqs.(\ref{NLs},\ref{NLi}), by letting 
 $\hat A_p (\w_p,z)  \to (2\pi)^{3/2} \alpha_p \delta(\w_p -\w_{0p})$,  where $\w_{0p} = (q_p, 0,0)$. Let us focus on a signal mode $\w_s$, for which 
\label{pareqs}
\begin{align}
\frac{\partial}{\partial z}   \hat{A}_s  (\w_s, z )   &=  g_0
\left[   \hat{A}_i^\dagger(\w_{0p} -\w_s -\Gx, z)  e^{-i D_1 (\w_s ) z }  +
\hat{A}_i^\dagger(\w_{0p} -\w_s +\Gx, z)  e^{-i D_2 (\w_s ) z }  \right]
\label{Ls}
\end{align}
where $g_0=\chi \alpha_p $ is taken real for definiteness,  and
\beq
D_{1,2} (\w_s) \equiv \DD_{1,2} (\w_s, \w_p= \w_{0p}) = k_{sz}(q_{sx}, q_{sy},\Omega_s)+k_{iz}(q_p - q_{sx} \mp  G_x, -q_{sy}, -\Omega_s )-k_{pz}(q_p,0,0)+G_z  
\label{DPW}
\eeq
are the QPM functions in the parametric limit.  This equation needs to be  coupled to the  evolution  of the two idler modes
\begin{align} 
\frac{\partial}{\partial z}   \hat{A}_i  (\w_{0p} -\w_s -\Gx, z )   &=   g_0
\left[   \hat{A}_s^\dagger( \w_s ,z)  e^{-i D_1 (\w_s ) z }  +
\hat{A}_s^\dagger(\w_s +2\Gx, z)  e^{-i D_2 (\w_s +2\Gx) z }  \right]
\label{Li1} \\
\frac{\partial}{\partial z}   \hat{A}_i  (\w_{0p} -\w_s +\Gx, z )   &=   g_0
\left[   \hat{A}_s^\dagger( \w_s  -2\Gx, z)  e^{-i D_1 (\w_s  -2\Gx ) z }  +
\hat{A}_s^\dagger(\w_s , z)  e^{-i D_2 (\w_s) z }  \right] 
\label{Li2} 
\end{align} 
%\esub
forming in principle an infinite chain of coupled equations. However,  in most cases, only one of the two nonlinear processes is effective, because for a given signal mode either $\D_1 (\w_s)=0$  or $\D_2(\w_s)=0$.  Then the usual pair of parametric equations coupling two signal-idler conjugate modes is obtained. Noticeably, there exist special points which are shared by both processes,  satisfying 
\beq
\D_1(\w_s)=\D_2(\w_s)=0 .
\label{SS}
\eeq
As it can be easily verified from Eq.\eqref{DPW},  the first equality requires that $q_{sx}=q_p$, i.e. all the  shared modes are characterized  by  the same  x-component of the wave-vector  as the pump. The y-component wave-vector then depends on the frequency $q_{sy}= q_{sy} (\Om_s)$, as determined by QPM (the second equality in Eq. \eqref{SS}).  Each  shared signal mode is then coupled to two idler modes at 
$q_{ix}= \mp G_x$, $q_{iy} =-q_{sy}$, $\Om_i= -\Om_s$ . 
\par 
The dual situation occurs for a shared idler mode at $q_{ix}= q_p$, coupled to two signal modes at $q_{sx}=\mp G_x$. Notice that unless the pump satisfies the {\em resonance} condition $q_p = \pm G_x$, the shared signal and shared idler configurations are strictly incompatible, and the evolution of each shared mode with its coupled modes forms a closed set of three parametric equations that will be examined in the next Sec.\ref{Sec:3mode}. Conversely, when the pump is tilted at $q_p = \pm G_x$, two triplets of 3-modes, initially uncoupled,  merge into a system of 4 coupled modes,  whose equations will be studied in Sec. \ref{Sec:4mode}.
%%%%%
\begin{figure}[h]
\includegraphics[scale=0.55]{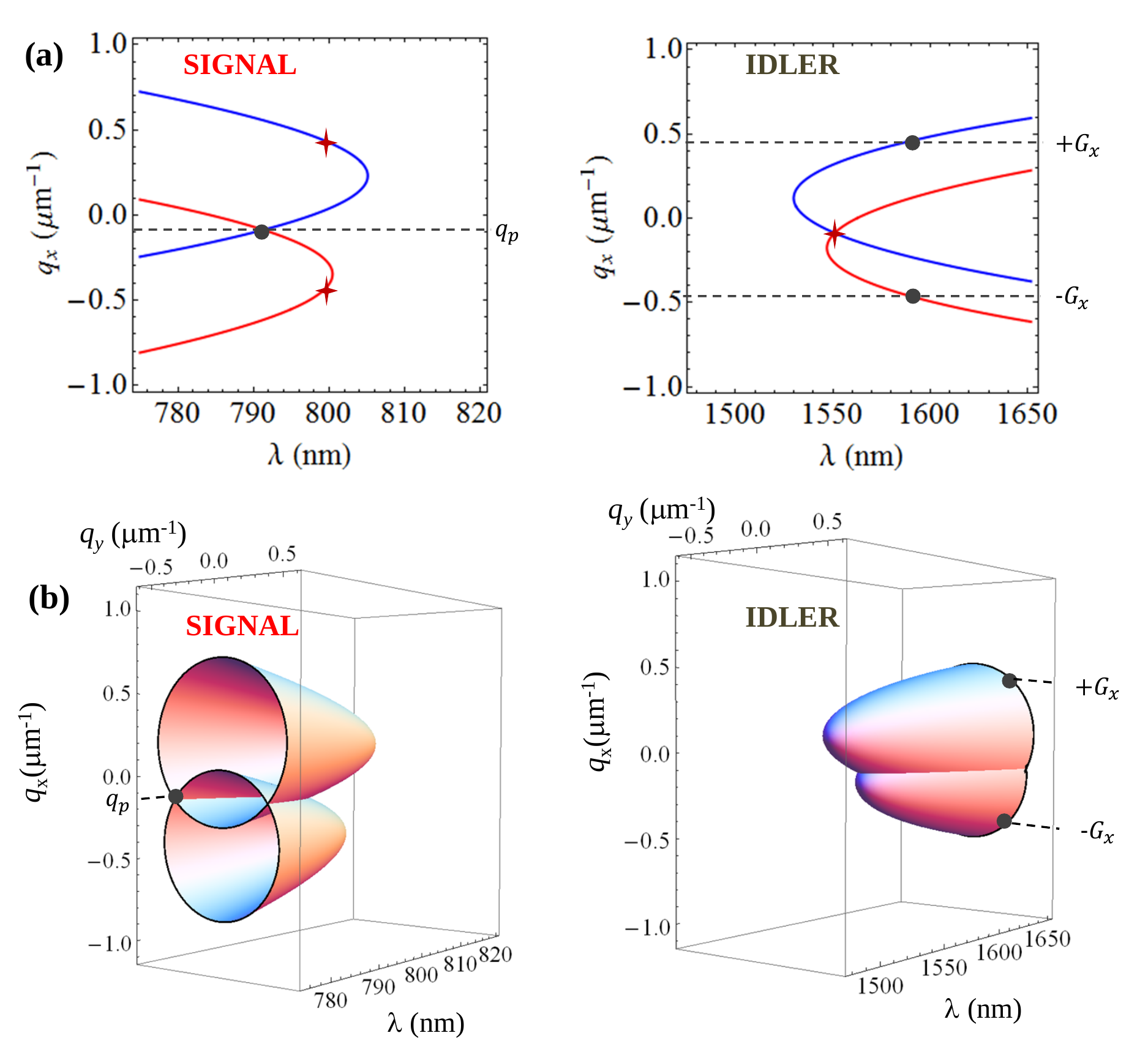}
\caption{ QPM surfaces in the Fourier space (see text),  calculated via the Sellmeier formulas in \cite{Lim2013}, for  $q_p= -0.3 G_x$  (non resonant pump).
%for the signal  (left) and  idler (right) fields. 
For better readability $\Om$ has been mapped to the wavelength $\lambda$. 
(a)  Projections along $q_y=0$ of the full 3D surfaces in (b).   The bullets mark on-axis (a) and off-axis (b) examples of the 3 entangled modes,  corresponding to a shared signal + 2 coupled idlers.  The stars show the dual configuration with a shared idler+ two coupled signals. Parameters are those of the 
non-degenerate HexNPC  crystal with $\Lambda= 8.3 \mu$m used in \cite{Gallo2018}.
 }
\label{fig2_QPM}
\end{figure}
%%%%%%%%%%%%%%%%
%%%%%
\begin{figure}[ht]
\includegraphics[scale=0.5]{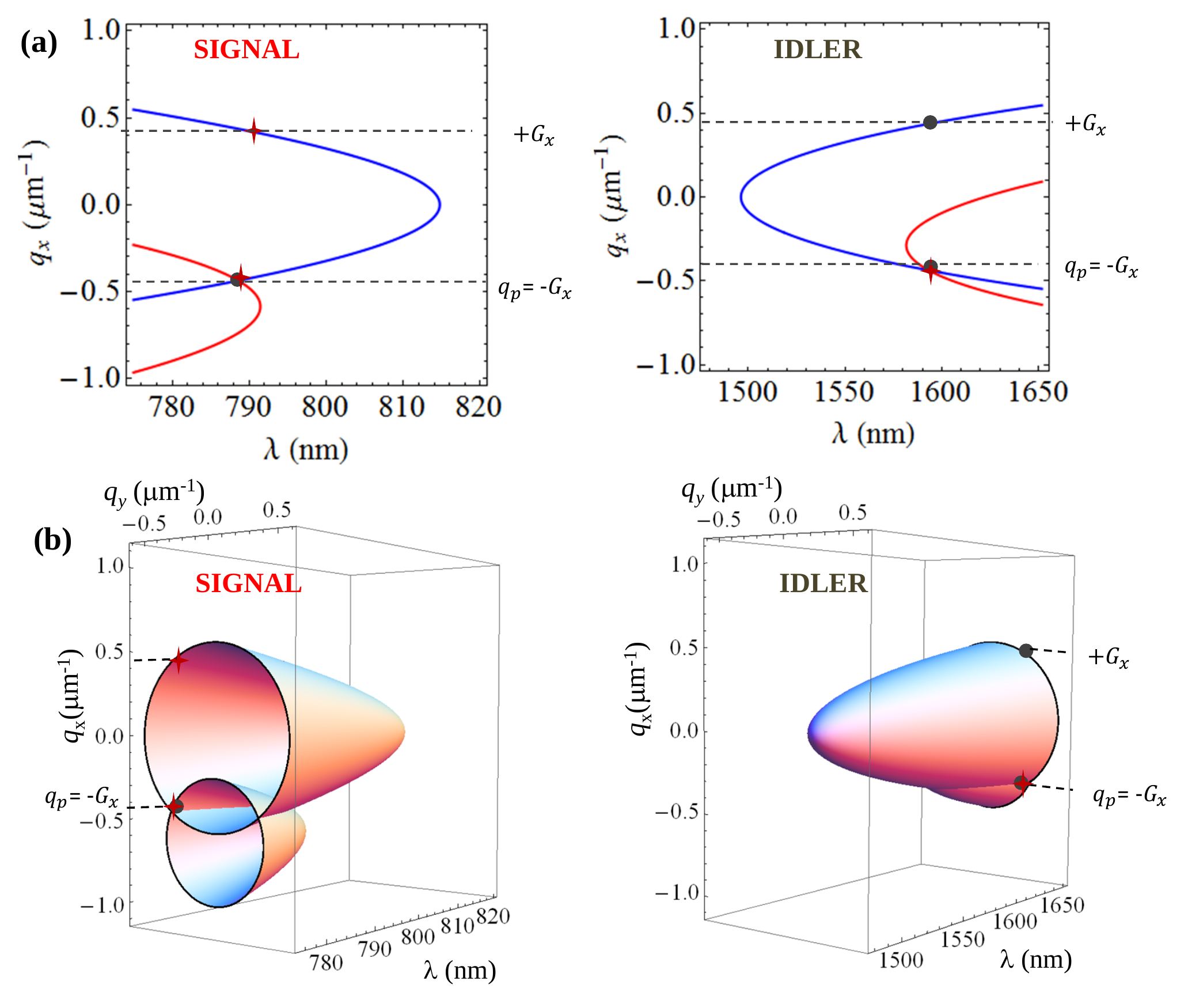}
\caption{Same as Fig.\ref{fig2_QPM} but at superresonance, with the pump tilted at 
 $q_p= - G_x$. In this condition,  all the shared modes superimpose to the lines of modes  at $q_x =-G_x$, and two triplets of hot-spots merge into  4 coupled modes, as observed  in \cite{Gallo2018}.}
\label{fig3_QPM}
\end{figure}
%%%%%%%%%%%%%%%%
\par
Figs.\ref{fig2_QPM} and \ref{fig3_QPM} show two examples of the QPM surfaces in the  Fourier space, away from superresonance and at superresonance, respectively.    In the signal panels, the lower and upper curves  show  the surfaces $\D_1(\w_s)=0 $ and  $\D_2(\w_s)=0$,   respectively,  with the shared modes lying at their intersections.  For the conjugate idler field,  the same surfaces are plotted as a function of $\w_i = -\w_s +  \w_{0p} 
 \mp \Gx$. 
The shared modes form continuous lines, which sit  on the plane $q_x= q_p$;  their coupled modes also form continuous  lines  sitting on the two planes $q_x = \mp G_x$. 
When the pump is tilted at $q_p= -G_x$,  all the shared modes superimpose to the lower coupled modes,  and a sudden increase in the intensity at the location of these modes occur, as demonstrated in \cite{Gallo2018}. 
%%%%%%%%%%%%%%%%%%%%%%%%
\section{3-mode entanglement} 
\label{Sec:3mode}
Let us  focus on the shared signal configuration: we will show that away from superresonance  this configuration realizes a 3-mode entangled state, which can be pictured as a  parametric process of gain $g=g_0 \sqrt{2}$ followed by a  beam splitter. 
The shared idler configuration can be treated in a completely analogous way.
\par
\begin{figure}[ht]
\includegraphics[scale=0.5]{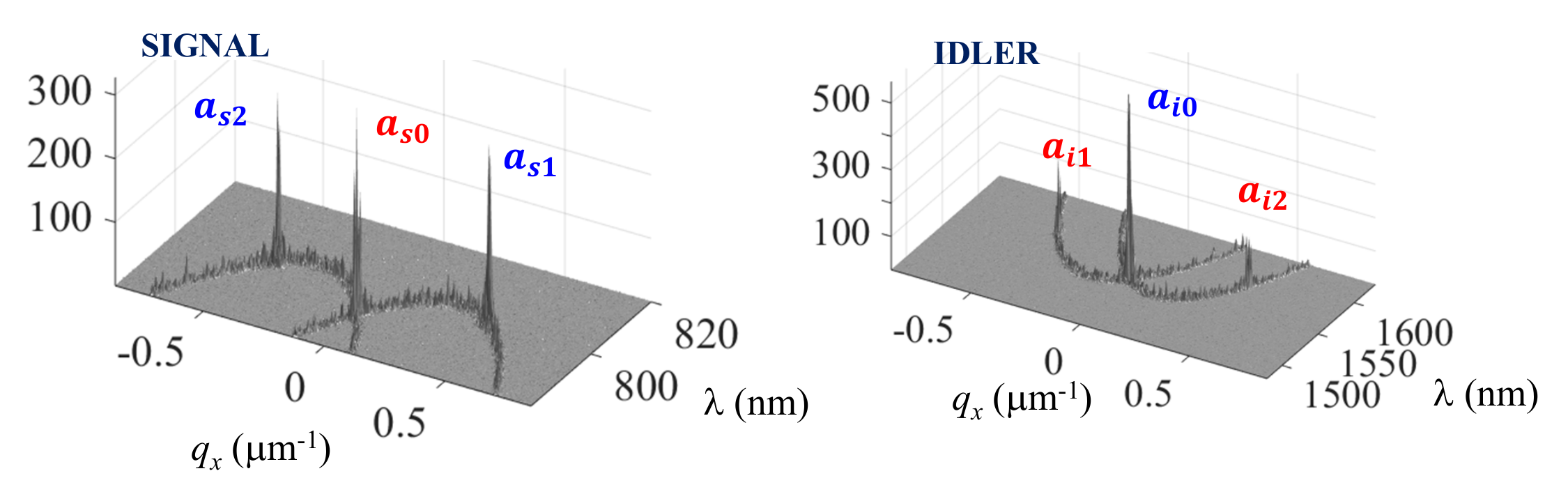}
\caption{Numerical simulations of the $(\lambda, q_x)$  distributions of the signal and idler photon-numbers  at the output of a  HexNPC  at $q_y=0$,  and for $q_p=0$ (non resonant pump).  Two triplets of hot spots corresponding to the shared signal and shared idler configurations are evident.  The pump is a 10ps Gaussian pulse centered at $\lambda_p=527.5$nm, with a spatial Gaussian profile,  of widths $600\,  \mu$m and $200 \, \mu$m in the x and y directions. The crystal length is $l_c= 10$mm,  and  $g_0 l_c= 5$.  }
\label{figHS}
\end{figure}

 
Let us consider a signal mode  $\w_s = (q_p,q_{sy},\Om_s)$ shared by both processes, i.e.  such that $\D_1(\w_s)= \D_2(\w_s): =\D(\w_s)  \approx 0$.  It can be easily shown that sufficiently away from the condition  $q_p =\mp G_x$,  the modes $\w_s \mp 2\Gx$ appearing in Eqs. \eqref{Li1} and \eqref{Li2} are not phase-matched, so that \eqref{pareqs} reduce to a closed set of three equations, coupling the shared signal with the two idlers at $q_x=\mp G_x$.  Let us give a short name to these modes 
\begin{align}
&\azero  := \As (q_p, q_{sy},\Om_s)  \qquad &\text {shared signal at $q_p$}\\
&\auno  := \Ai (-G_x, -q_{sy},-\Om_s )  \qquad &\text {coupled idler  at $-G_x$}\\
&\adue := \Ai (+G_x, -q_{sy},-\Om_s )  \qquad &\text {coupled idler  at $+G_x$}
\end{align}
Their evolution along the sample is described by 
\bsub
\label{ssprop}
\begin{align}
&\frac{\partial}{\partial z}   \azero  (z) =  g_0 
\left[  \auno^\dagger   (z)
 +    \adue^\dagger (z)  \right] e^{-i \D (\w_s)z }
\label{sprop}\\
&\frac{\partial}{\partial z}   \auno     (z)=  g_0 
\azero^\dagger  (z)  e^{-i \D ( \w_s) z } 
\label{i1prop} \\
&\frac{\partial}{\partial z}   \adue   (z)  =  g_0  \azero^\dagger   (z)e^{-i \D( \w_s) z } 
\label{i2prop}
\end{align}
\esub
These equations can be easily solved by introducing  the canonical transformation 
$
\ah_{i\pm}  =  \frac{ \auno  \pm  \adue   }{\sqrt{2}}  \, $,
%\label{trf}
%\qquad  &\text {"cosine"  mode}\\
%\ah_{i-} (-\w_s) &=  \frac{ \ai  (-\w_s  ) - \ai  (-\w_s ) }{\sqrt{2}}  \qquad  &\text {"sine"  mode}
%\end{align}
%which is a canonical tranformation that preserves commutation relations, so that $\ah_{i+}$ and $\ah_{i- }$ are two independent photon anihilation operators. The transformation \eqref{trf} is equivalent to the action of a 50:50 non symmetrical beam-splitter, characterized by a transmission coefficient  $t=1/\sqrt{2} $ and  reflection coefficients  $ r= \pm 1/\sqrt{2}$ on its two faces, acting on  the two tilted  idler modes  at  $q_{ix} =\pm G_x$.  
leading to: 
\bsub
\label{ssprop2}
\begin{align}
&\frac{\partial}{\partial z}   \azero (z )   =   \sqrt{2}g_0
  \ah_{i+}^\dagger  (z) 
 e^{-i \DD (\w_s) z }
\label{sprop1}\\
&\frac{\partial}{\partial z}   \ah_{i+}  (z)   =   \sqrt{2}g_0
\azero^\dagger  (z)   e^{-i \DD ( \w_s) z } 
\label{i+prop} \\
&\frac{\partial}{\partial z}    \ah_{i-}  (z)  =  0
\label{i-prop}
\end{align}
\esub
The last equation simply 
%\eqref{i-prop}
 means that  the difference mode $  \ah_{i-} $  is not affected by the parametric process:  if  at the crystal entrance face the idler field is in the vacuum state, then the difference between any two symmetrical idler modes at $q_{ix} = \mp G_x$ remains in the vacuum.  However,  each of  the two idler output  modes  has a nonzero intensity, because it has been parametrically amplified: as we shall see, this implies a  correlation between the two  idlers coupled via the same shared signal. 
 The first two equations are the usual coupled 
 equations, describing parametric generation for a pair of signal-idler conjugate modes, with an enhanced parametric gain 
\beq
g = \sqrt 2 \,  g_0
\label{gsqrt2}
\eeq
As first shown in \cite{Liu2008} and then demonstrated in \cite{Levenius2012, Chen2014, Gallo2018},   in the high-gain regime $g_0 l_c \gg 1$,  
%where the intensity grows exponentially with the propagation length,   
this local gain enhancement gives rise to bright hot-spots in the parametric emission at the location of the three modes (see Fig.\ref{figHS} and \ref{figqxqy}) . There,  
$ \langle \hat a_j ^\dagger \hat a_j \rangle \propto \sinh^2 (\sqrt 2 g_0  z) \simeq  \left(e^{2g_0 z} \right)^{\sqrt 2} $,  $(j=s0, i1,i2)$, and  the  increase of intensity with respect to the background 2-mode fluorescence  follows a power law $ I_{3-mode}  = (I_{background} )^{\sqrt 2}$ \cite{Gallo2018}. 
%%%%%
%%%%%%%%%%%%%%%%
Notice  that in the direct space  the sum  and difference modes have  spatial modulations  $\sim \cos (G_x x)$, and $\sim \sin (G_x x)$, respectively, in-phase and out-of-phase with   the transverse modulation of the nonlinearity. Thus, the enhanced gain of the $\apiu$ mode can be interpreted as a spatial resonance with the nonlinear lattice. 
\par
\begin{figure}
{\scalebox{.5}{\includegraphics*{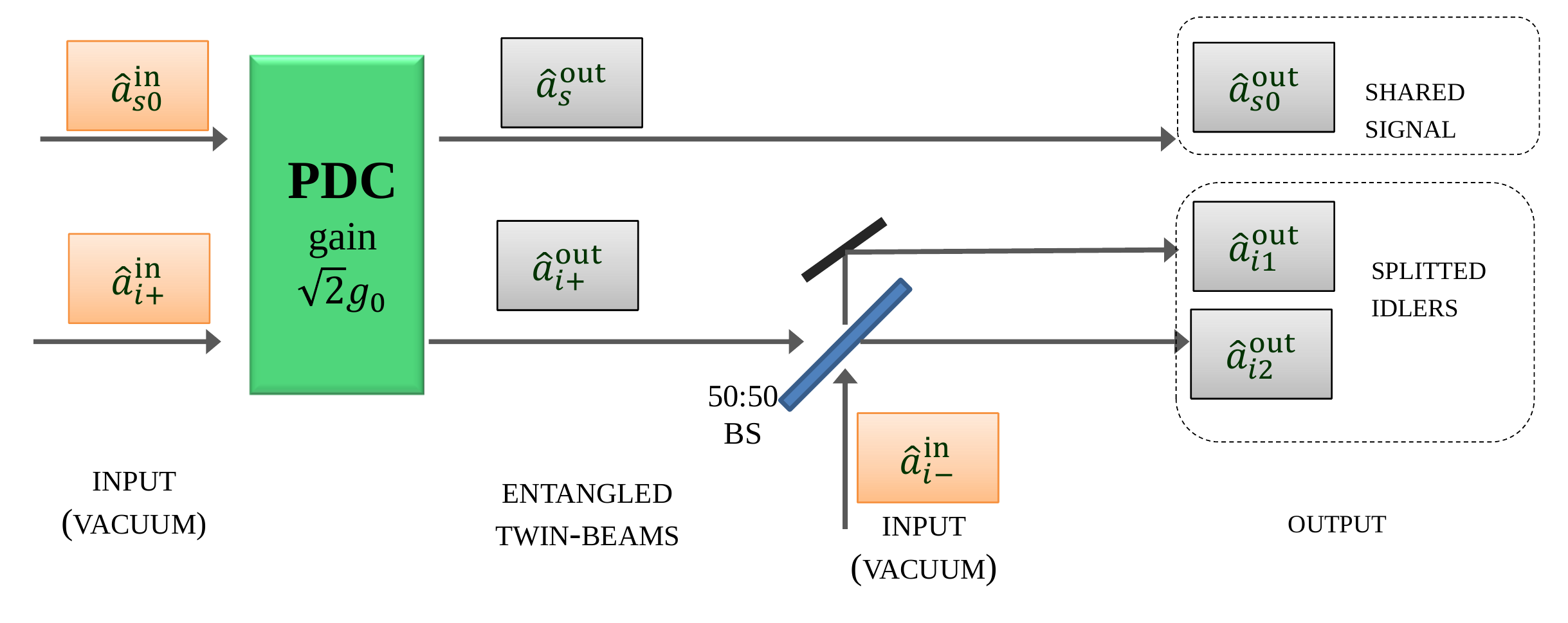}}}
\caption{Unfolding  of  the  3-mode entanglement generated by the HexNPC away from resonance,  where a  shared signal  mode $\azero$ is coupled to two  idlers $\auno$ and $\adue $.
 The process is equivalent to the a single PDC process  generating  maximally entangled twin beams, followed by the action of a 50:50 beam-splitter that mixes one of the entangled  twins  with an independent  vacuum field (the difference between the two input idlers). As a result, tripartite entanglement  is realized, with the  two output idlers being entangled, but not perfectly, to the signal, and being correlated one to each other. }
\label{fig_solution}
\end{figure}
\begin{figure}[h]
\includegraphics[scale=0.65]{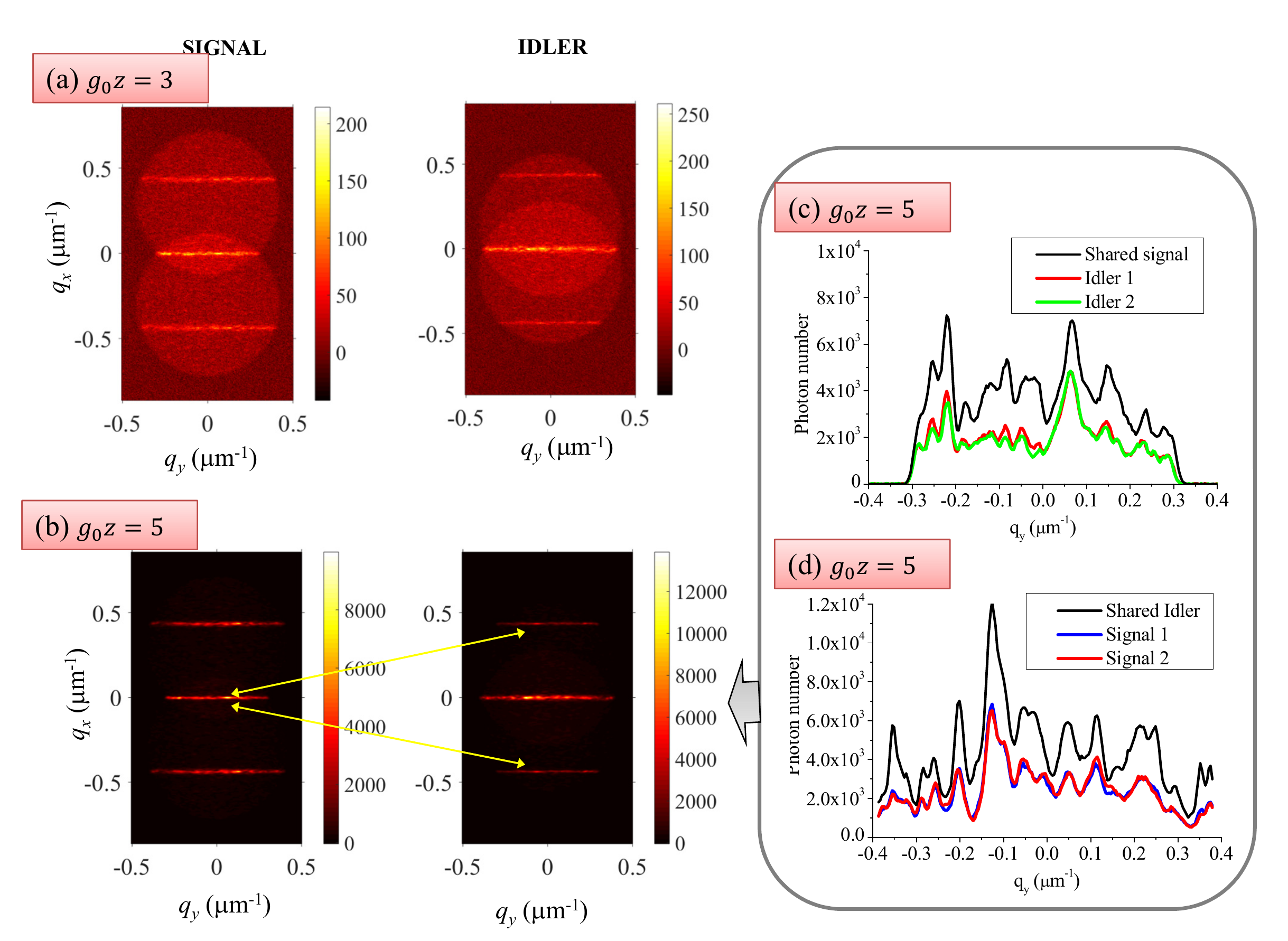}
\caption{ (a) and (b) Numerical simulations of the$(q_x, q_y)$  photon-number distributions (integrated over a single pump shot),   for $q_p=0$, showing the emergence of hot-spot lines for increasing propagation distance.  Shared modes  are located at $q_x=q_p=0$,  while their coupled modes are at  $q_x = \mp G_x$ . (c)  Sections along $q_y$, highlighting  a spatial correlation among the intensity fluctuations of the three modes  ($\azero, \auno,\adue$). Notice that the idler distributions have been mirrored $q_y \to -q_y$ . (d) Same as (c) for the shared idler configuration. Other parameters as in Fig. \ref{figHS}. }
\label{figqxqy}
\end{figure}

Coming to the quantum properties of the state,   the explicit solution of the two coupled  parametric equations \eqref{sprop1} and \eqref{i+prop} can be found within the  standard input-output formalism (see e.g. \cite{Gatti2017} for a summary), as a  Bogoliubov transformation linking the quantum operators at the crystal exit face  $\hat a_j^{ \mathrm out} = \hat a_j  (l_c)$  to those at the input $\hat a_j^{ \mathrm in} = \hat a_j  (0)$,  $(j=s0, i+)$:
\bsub
\label{inout1}
\begin{align}
& \asout = \Ug (\w_s) \, \asin   + \Vg (\w_s) \, \apiu^{\dagger \, in}
\label{inouta}\\
& \apiuout  = \Ug (\w_s) \, \apiuin + \Vg (\w_s) \, \as^{\dagger \, in}
\label{inoutb}
\end{align}
 Here $\gamma= \sqrt{2}= g/g_0$ is a parameter giving the local gain enhancement in the hot-spots, and
the explicit expressions of $\Ug$ and $\Vg$ can be found in Appendix \ref{AppA}. 
A well known consequence of this transformation (see e.g. \cite{Knight2005a}) is that   $\ah_{i+}$  is the twin beam of $\azero$, and their joint state is the {\em two-mode squeezed vacuum}.  In the continuous-variable  domain, the usual picture of twin-beam entanglement holds:    the intensity fluctuations of $\ah_{i+}$ and $\azero$ are perfectly correlated, and there is  a "Einsten-Podolsky-Rosen" \cite{Einstein1935,Reid1989} correlation between their field quadratures. Equations \eqref{inout1} have  to be considered together with 
\beq
\amenout =\amenin 
\label{inout2}
\eeq
and the back-transformation 
\begin{align}
&\hat a_{i1,2}=  \frac{ \apiuout\pm\amenout  }{\sqrt{2}}\, .
%= \frac{ \apiuout(\w) +\amenin (\w) }{\sqrt{2}} \\
%&\adueout (\w) = \frac{ \apiuout(\w) -\amenout (\w) }{\sqrt{2}} 
%=\frac{ \apiuout(\w) -\amenin(\w) }{\sqrt{2}}
\label{back1}
\end{align}
\esub
Equations \eqref{inout1} are sufficient to calculate all the quantities of interest starting e.g. from a vacuum input. However, most of the properties of the 3-mode entanglement can be understood by noticing that the transformation  \eqref{back1} can be formally described by the action of a 50:50 beam-splitter, mixing $\apiuout $, which is the twin beam of the shared signal $\asout$, with 
$\amenout$, which trivially coincides with 
$\amenin$. 
The overall process can be schematically pictured (Fig \ref{fig_solution}) as 
\begin{itemize}
\item The action of a PDC device, generating the  entangled twin beams $\asout$, $\apiuout$  with parametric gain $g= \sqrt 2 g_0
$.
\item  Followed by a 50:50 beam-splitter acting on the idler arm, that mixes $\apiuout $ with an independent mode $\amenin$.
\end{itemize}
All of this is implemented by the nonlinear photonic crystal, which can be viewed as  a sort of monolithic  3-mode nonlinear interferometer. 
%%%%%%%%%%%%%%%%%%%%%%%%%%%
%%%%%%%%%%%%%%%%%
\par 
Clearly, each of the two splitted idlers is quantum entangled to the signal, even though the entanglement is not the maximal one of the twin beam, because of  the random vacuum fluctuations of  $\amenin$ entering the other port of the beam-splitter (assuming the difference mode is in a vaccum or coherent state). In addition, the coupling via the shared idler induces a certain degree of correlation between the splitted idler beams. Precisely : 
\begin{itemize}
\item
 If the signal is {\em undetected}, the correlation between $\aunoout$  and  $\adueout$  is classical, and is equivalent to the correlation between  the two outputs of a beam splitter illuminated by a thermal light beam. This follows straightforwardly from the well known fact that the marginal  statistics of each twin beam, when considered independently from the other,  is  thermal-like. Then, in the high-gain  the two idlers may possess  a  very high degree of mutual coherence, just as the splitted thermal beams used for thermal ghost imaging \cite{Gatti2004, Gatti2006}, but their correlation is always shot-noise limited. 
\item
If the signal is {\em detected}, the statistics of the coupled idlers, conditioned to detection (of some light observable) in the signal arm can be nonclassical, in particular they may show anticorrellated intensity fluctuations. 
% A simple example is the following: in the low-gain regime, assume to detect a single photon in the signal arm:then the  state of the coupled idlers,  conditioned to this detection,  correspond to a single photon state (Fock state) divided by a beam splitter which is  perfectly anticorrelated , because the idler photon is either in arm 1 or 2, never in both
\end{itemize}
To clarify this last point, the problem can be reformulated in  terms of the quantum state of the system. 
 Let us focus on a specific shared-signal mode $\w$, with its coupled idlers (to be precise, one should also perform a discretization of  Fourier modes, but it really does not matter). Given the input-output relations \eqref{inout1}  for  the transformed modes $ \azero, \apiu, \ameno$,, their output state can be written as: 
\begin{align}
| \psi ^\prime_{out}\rangle& = \hat R_{s+} (\xi) \hat {\text{1}}_-  | \psi _{in}\rangle
\end{align} 
where  the suffixes "$s,+,-$ " refer to the shared signal ,  to  the sum of  the two idlers and to their  difference, respectively, $\hat {\text{1}}_-$ is the identity operator, and 
$\hat R_{s+} (\xi) = \exp {\left( \xi \azero^\dagger \hat a^\dagger_{i+} -\xi^\star \azero \apiu\right)  } $,  is the {\em two-mode squeeze operator}, acting on the shared signal and the $"+"$ mode.  The squeeze parameter $\xi $  can be related to the  input-output coefficients in Eq.\eqref{inout1}  (appendix \ref{AppA}),  and for phase-matched modes  reduces to  $ \xi = \sqrt 2  g_0 l_c$. 
%$\tanh (|\xi_\eta|)= \frac{V_\eta (\w)}{U_\eta(\w)}$
Provided that the input state is vacuum, then
\begin{align}
| \psi^\prime _{out}\rangle& = \left[ \sum_{N=0}^{\infty}  c_N |N\rangle_s \, |N\rangle_+\right] \otimes   |0\rangle_- 
%c_N &= \frac{ \left[ U(\w) V(\w) \right]^N}{ \left| U (\w) \right|^{2N+1} }
\end{align}
where  $|N\rangle $ denotes the  Fock state with N photons, $|0\rangle$ is the vacuum state, and $c_N = \frac{ \left[ \Ug(\w) \Vg(\w) \right]^N}{ \left| \Ug (\w) \right|^{2N+1} }$, ($\gamma = \sqrt 2$). Incidentally, one can readily  see that the reduced state in  the idler arm  is the thermal-like state: 
$\hat \rho_{\pm} = {\mathrm Tr}_s \left\{ | \psi^\prime _{out}\rangle\, \langle \psi^\prime _{out} \right\}
= \left[\sum_{N=0}^{\infty}  | c_N |^2  |N\rangle_+  \phantom{i}_+\langle N| \right] \otimes   |0\rangle_- \phantom{i}_- \langle 0 | $.\\
% The state inside square brackets   is the {\em  two-mode squeezed state}  of the twin beams,  characterized by having only terms with the same number of photons in the two modes.  The state of mode "-" is simply the vacuum state, because the idler difference does not evolve along the slab. 
For the original modes  ($\azero,\auno,\adue$),  the state is then obtained as $| \psi _{out}\rangle = \hat B_{+-}\, | \psi^\prime _{out}\rangle$, where $\hat B_{+-} $ is the generator of the beam-splitter transformation $   \hat B_{+-} \hat a_{\pm}   \hat B_{+-} ^\dagger = 
  \frac{ \auno \pm \adue } {\sqrt{2}} $.
By using then  standard properties of Fock states, 
\begin{align}
| \psi _{out}\rangle
%& = \hat B^{dagger}{50:50} \sum_{N=0}^{\infty}  c_N  |N\rangle_s    \otimes  \frac{ (\apiu^{\dagger} )^N}{\sqrt{N!}}     \, |0\rangle_+ \,  |0\rangle_- \\
&= \sum_{N=0}^{\infty}   c_N  |N\rangle_s   \otimes  \frac{1}{  \sqrt{N!}}  
\left(  \frac{ \auno^{\dagger} + \adue^{\dagger} } {\sqrt{2}}   \right)^N    |0\rangle_1 \, |0\rangle_2 \\
%&= \sum_{N=0}^{\infty}  c_N  |N\rangle_s    \otimes  \left[   \frac{1}{ \sqrt{N!}}   \frac{1}{(\sqrt{2})^N}
%\sum_{k=0}^{N}         (\auno^\dagger)^k \adue^{N-k} \frac{N!}{k! (N-k)!}  b|0\rangle_1 \, |0\rangle_2  \right]\\
&= \sum_{N=0}^{\infty}  c_N    |N\rangle_s  \left[ 
\sum_{k=0}^{N}       \frac{1}{(\sqrt{2})^N}     \sqrt{\frac{N!}{k! (N-k)!}  }
  |k \rangle_1 \, |N-k \rangle_2 
\right] 
\label{state_fin}
\end{align}
where  the binomial expansion formula has been used to step to the second line.  The result \eqref{state_fin} is the sum of infinite terms: each of them  is the product tensor of a state with  N photons in the signal mode, times the superposition of  all the possible partitions  of the  N twin idler photons  into k photons in mode 1 and N-k photons in mode 2. 
Assuming that {\em exactly N photons} are detected in the signal arm  the state of the two idler modes conditioned to this measurement is
\beq
| \psi _{idler}/_\text{N signal photon}\rangle = \frac{ \phantom{}_s\langle N | \psi_{out} \rangle} { || \phantom{}_s\langle N | \psi_{out} \rangle|| } = 
\sum_{k=0}^{N}       \frac{1}{(\sqrt{2})^N}     \sqrt{   \frac{N!}{k! (N-k)!}  }
  |k \rangle_1 \, |N-k \rangle_2 
\eeq
This state is the N-photon path-entangled state, corresponding to the superposition of all the possible partitions of N photons into  k photons in arm 1 and  N-k photons in arm 2, with probability  
$
P_{k,N-k}= \frac{1}{2^N}  \frac{N!}{k! (N-k)!}
$
following the  binomial distribution
($\frac{1}{2}$  being the probability of taking either path for each photon).  Such a state simply describes the random partition of the N photons in the two idler arms  with equal probability of  going one way or the other. 
It can be easily shown that the state is antibunched, and exhibits the maximum level of anticorrelation in the photon number fluctuations allowed by quantum mechanics.  The simplest example is for $N=1$, 
\beq
\left | \psi_{idler} / (\text {1 signal photon} )  \right\rangle =  \frac{1}{\sqrt{2}} \left( | 0 \rangle_1 \, | 1\rangle_2 + | 1 \rangle_1 \, | 0\rangle_2  \right)
\eeq
which is the path entangled single photon state  explored in \cite{Jin2013}. Conversely, when the signal is not detected, the state of the two splitted idlers is non-entangled because it is obtained by a linear transformation acting on the classical thermal state (but can be highly correlated and have mutual phase coherence \cite{Gatti2004b,Gatti2006}). 
%%%%%
%%%%%%%%%%%%%%%%
\par
We remind that in a full view of the problem, this 3-mode entangled state can be generated  for a continuous range of  Fourier modes $w_s$ satisfying  the shared signal condition \eqref{SS}, as well as for their shared idler counterparts.  These modes span a broad band of frequencies, as the $q_y$ coordinate is varied, or in practice, as the y-direction is scanned in the far-field (see \cite{Gallo2018} Supplementary Information). 
%%%%%%%%%%%%%%%%%%%%%%%%%%%%%%%%%%%%%%%%%%%%%%%%%%%%%%%%%%%%%%%%%%%
\section{Golden Ratio entanglement} 
\label{Sec:4mode}
%%%%%
In this section we focus on a particular pump  propagation direction,  corresponding  to a transverse spatial  resonance between the pump and  the nonlinear lattice. We demonstrate in this condition the emergence of a peculiar 4-mode entangled state, dominated   by the {\em Golden Ratio} of the segment. 
\par
\begin{figure}[h]
\includegraphics[scale=0.65]{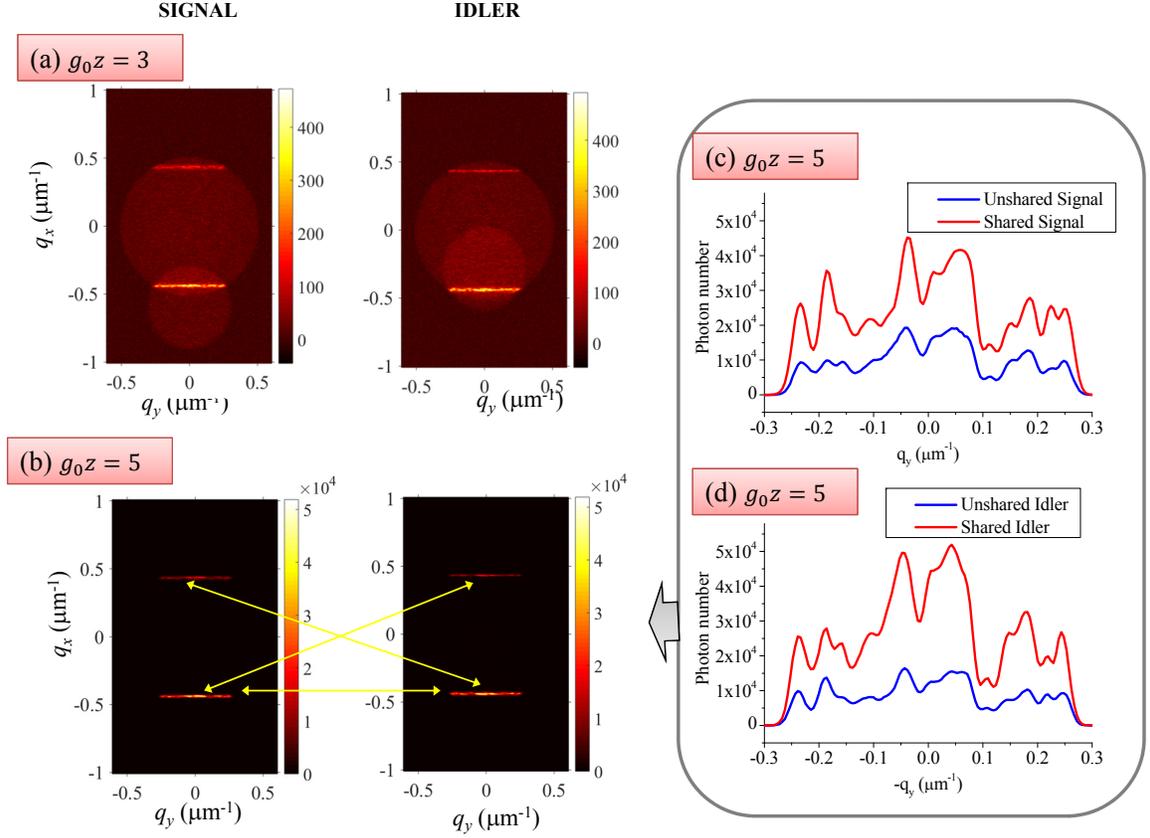}
\caption{Pump tilted at resonance $q_p=-G_x$. (a) and (b) Photon-number distributions  in the $(q_x, q_y)$  plane,   at the output of a nondegenerate HexNPC, showing the emergence of 4 lines of hot-spots. Those at $q_x=-G_x$  are shared by both processes (modes $\hat b_s$ and $\hat b_i$), while  those at $q_x = + G_x$ (modes $\hat c_s$ and $\hat c_i$) are unshared. The yellow arrows schematically show  the  coupling among the 4-modes. (c)  Sections along $q_y$ of the signal hot spots.  (d) Same for the idler but mirrored $q_y \to -q_y$ , in order to evidence the correlation. Other parameters as in Fig.\ref{figHS}.  }
\label{figresonance}
\end{figure}
\begin{figure}[t]
\includegraphics[scale=0.45]{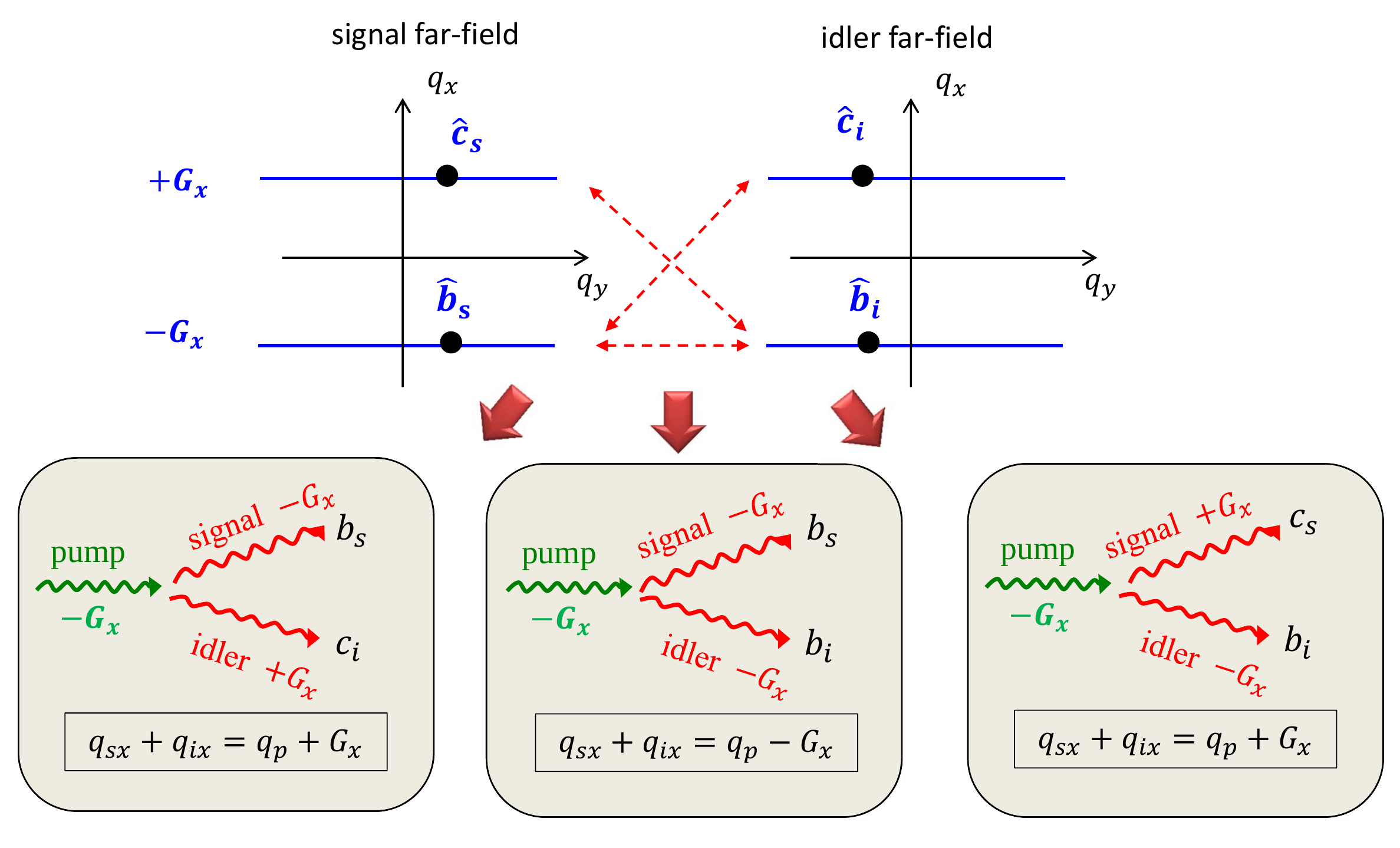}
\caption{Scheme of the coupling between the 4 modes at resonance, leading to Eqs.\eqref{resonance_prop}, and of the 3 microscopic down-conversion processes. The shared modes $\hat b_j$ are populated by two  processes out of the three. In the spontaneous regime, modes $\hat b_j$  are twice as intense as the unshared modes $\hat c_j$, while in the stimulated regime the ratio between the two intensities is $\phi^2$.  }
\label{figinteractions}
\end{figure}
\begin{figure}[ht]
\includegraphics[scale=0.5]{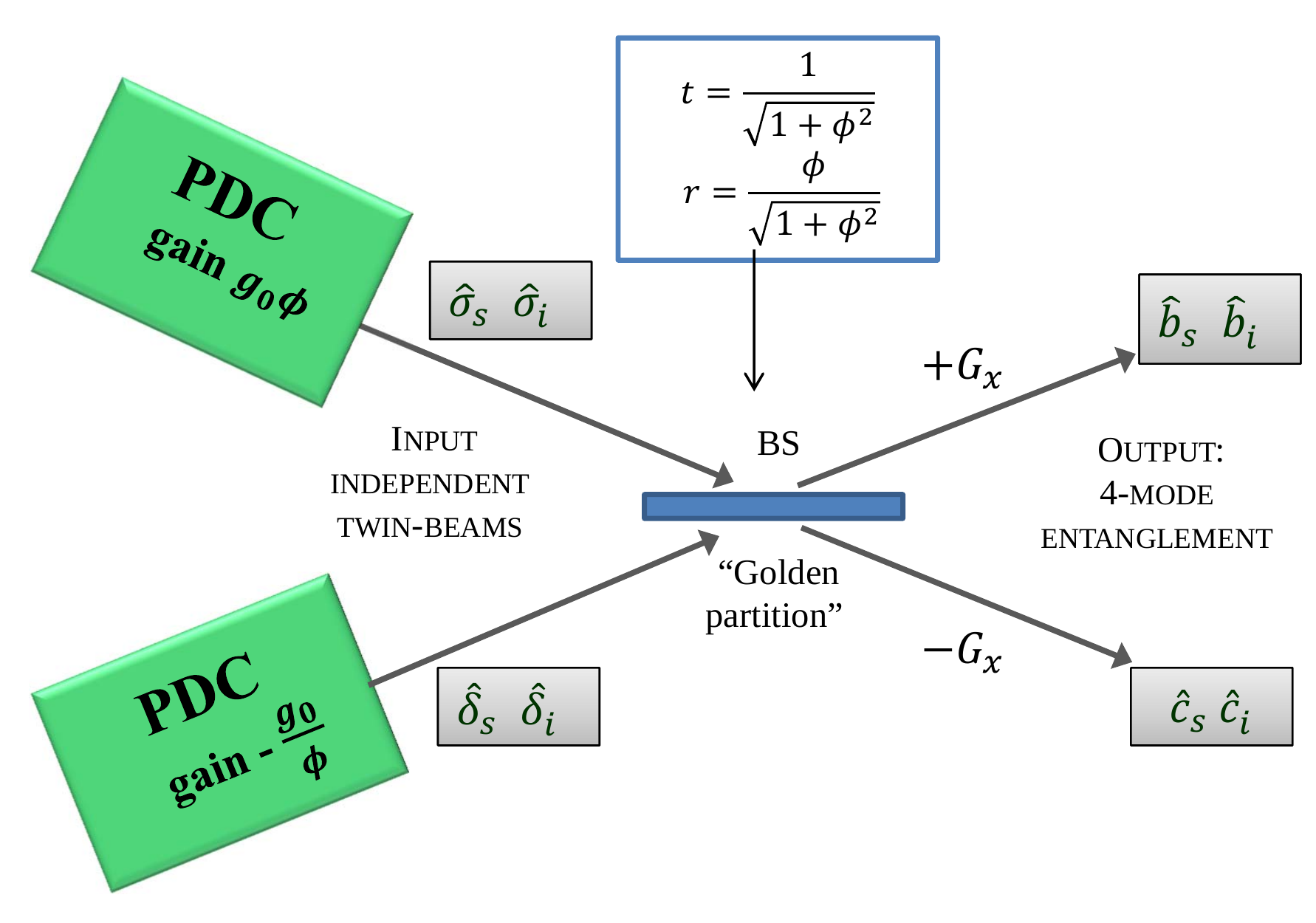}
\caption{Unfolding of the 4-mode entanglement generated  in the hexagonal nonlinear photonic crystal at pump resonance. This can be considered equivalent to two independent parametric processes with gains $g_0 \phi$ and $-g_0/\phi$, mixed on a beam-splitter which makes a golden partition of the two pairs of twin beams into 4 entangled modes. 
 }
\label{figgolden}
\end{figure}

By tilting the pump direction away from the symmetry $z$-axis,  the position in the far-field plane  of the shared modes move  along $q_x$ jointly with $q_p$,  while the position of the coupled modes remain  fixed at $q_x=\pm G_x$.  Thus,  when  $q_p = \mp G_x $,  in each signal/idler beam the shared modes arrive to superimpose to one line of unshared mode to which they were originally uncoupled (Fig. \ref{fig3_QPM}). As a result, the three hot-spot lines of the signal and the idler far-fields degenerate into two lines symmetrically positioned at $q_x=\pm G_x$ (Fig.\ref{figresonance}), with a sudden increase of their intensity \cite{Gallo2018}.  
%%%%%%%%%%%%%%%%%%%%%%%%%%%
%%%%%%%%%
\\
We can describe this phenomenon as a  {\em spatial resonance} between the pump and the lattice, because 
  the pump has a transverse modulation (in its phase, not in its intensity) with the same spatial periodicity $\Lambda_x= \frac{2\pi}{G_x}$ as the nonlinear grating. Notice that in this conditions he same  modulation  characterizes  all the hot-spots of the down-converted beams.
%so that we can talk of a {\em superresonance} effect. 
%so that we can talk of a { resonance  effect}  in its classical sense, because the three interacting fields have the same x-periodicity of the nonlinear grating.  
\\
Let us assume for definiteness $q_p = - G_x$,  so that  shared and coupled modes    merge  at $q_x=-G_x$.  If we focus on a specific y-direction  and on pair of conjugate signal idler frequencies   (Fig.\ref{figresonance}),  two triplets of hot spots which were originally uncoupled coalesce 
into 4 coupled modes. Let us give a name to these modes, as in the scheme of Fig. \ref{figinteractions}: 
\begin{align}
&\bs := \As (-G_x, q_{sy},\Om_s) & \bi  :=  \Ai (-G_x, -q_{sy},- \Om_s)\qquad &\text {shared modes at $-G_x$}\\
&\cs  := \As (+G_x, q_{sy}, \Om_s )  & \ci  :=  \Ai (+G_x, -q_{sy},- \Om_s)\qquad &\text {unshared modes  at $+G_x$}
\end{align}
%%%%%%%%%%%%%%%%%%%%%
%%%%%%%
Denoting by  $\w_s = (-G_x, q_{sy}, \Omega_s) $  the coordinate of the shared signal, the equations describing their  evolution along the sample take  the closed  form 
\bsub
\label{resonance_prop}
\begin{align}
&\frac{\partial}{\partial z}   \bs  (z)   =  g_0
\left[  \bi^\dagger  (z) 
 +     \ci^\dagger(z)   \right] e^{-i \D (\w_s) z }
\label{bsprop}\\
&\frac{\partial}{\partial z}   \cs  (z)   =  g_0\bi^\dagger(z)  e^{-i \D (\w_s) z } \label{csprop}  \\
&\frac{\partial}{\partial z}   \bi  (z )   =  g_0
\left[  \bs^\dagger(z) 
 +     \cs^\dagger(z)   \right] e^{-i \D (\w_s) z }
\label{biprop}\\
&\frac{\partial}{\partial z}  \ci  (z )   =  g_0\bs^\dagger(z)  e^{-i \D (\w_s) z }
\label{ciprop}
\end{align}
\esub 
 It is not difficult  to find their solutions  in the spontaneous regime, that corresponds to the limit $g_0 l_c \ll 1$.  As we shall see in the following,  the mean number of photons in each of the 4 modes, at the leading order in $g _0 z$ read
\begin{align}
\langle  \bs^\dagger (z)  \bs ( z)\rangle &= \langle  \bi^\dagger ( z)  \bi (z)\rangle =2 (g_0 z) ^2    \sinc ^2 \left( \frac{\D (\w_s) z}{2}\right)  \delta (0)
\\
\langle  \cs^\dagger (z)  \cs (z)\rangle &=  \langle  \ci^\dagger ( z)  \ci ( z)\rangle
=
(g_0 z)^2  \sinc^2 \left( \frac{\D (\w_s) z}{2}\right)  \delta (0)
\label{int_low}
\end{align}
where  the $\delta (0) $ is an artificial divergence coming from the plane-wave pump approximation which can be easily
 removed\footnote{the $\delta (0) $ will then be replaced by a term $ \propto \frac {1}{\Delta \Omega_p \, \Delta q_p^2} $, where $\Delta \Omega_p, \Delta q_p$ are the temporal and spatial frequency bandwidths of the pump}.  Thus,  we see that in the spontaneous regime modes $\hat b_j$ are twice as populated  as modes $\hat c_j$. This  is natural because out of  the three microscopic processes allowed by energy-momentum conservation,  two of them contribute to the  population of  e.g  the shared signal mode $\hat b_s$,  and only one process contributes to the population of the unshared mode $\cs$,  as shown  in Fig. \ref{figinteractions}b. 
\\
The situations changes in the  stimulated regime of PDC. Indeed,  the linear system \eqref{resonance_prop} can be solved for any value of $g_0$, by introducing 
 the transformation 
%
%\begin{align}
%\begin{cases} 
%\hat \sigma_j = \frac{ \phi \hat b_j+ \hat c_j } {\sqrt{1+\phi^2}} \\
%\hat \delta_j  = \frac{ \hat b_j  - \phi \hat c_j } {\sqrt{1+\phi^2}} 
%\end{cases}  \qquad (j=s,i) 
%\label{trphi}
%\end{align} 
%%%%
\begin{align}
\begin{pmatrix} 
\hat \delta_j \\[0.8em]
\hat \sigma_j
\end{pmatrix} 
=  
%\frac{ 1}{\sqrt{1+\phi^2}} 
\begin{pmatrix}
\frac{ 1}{\sqrt{1+\phi^2}} \; &  \frac{ -\phi}{\sqrt{1+\phi^2}}  \\[0.8em]
\frac{ \phi}{\sqrt{1+\phi^2}} \; &  \frac{ 1}{\sqrt{1+\phi^2}}
\end{pmatrix} 
\begin{pmatrix} 
\hat b_j \\[0.8em]
\hat c_j
\end{pmatrix} 
\qquad (j=s,i) 
\label{trphi}
\end{align} 
where $\phi$  is the so called {\em Golden Ratio } of the segment
\beq
\phi= \frac{1+ \sqrt{5} }{2} = 1.618..
\eeq
This sort of magic irrational number \cite{Dunlap2017} is the solution to the problem of partitioning a segment into two parts $b$ and $c$, whose ratio is equal to the ratio between the total length of the segment and the longer part: 
$
\frac{b}{c}= \frac{b+c}{b}= \phi
$.
 The Golden ratio can be also put in relation with the Fibonacci sequence (see e.g \cite{Koshy2011ch2}, \cite{Dunlap2017}), and is the asymptotic value to which the ratio between two consecutive numbers in the sequence converge. 
With few steps which makes use of the properties of the Golden ratio,  the original 4-mode equations can be decoupled  into two independent systems 
%\bsub
%\label{sigma_prop}
\begin{align}
\label{sigmaprop}
\begin{cases} 
\frac{\partial}{\partial z}   \hat \sigma_s  (z )   =  g_0 \phi
 \hat \sigma_i^\dagger( z)  e^{-i \D (\w_s)z }
%\label{sigmasprop}\\
\\[0.7em]
\frac{\partial}{\partial z}   \hat \sigma_i  ( z )   =  g_0 \phi
 \hat \sigma_s^\dagger (z)  e^{-i \D(\w_s) z }
\end{cases}
\end{align}
%\esub
and 
\begin{align}
\label{deltaprop}
\begin{cases} 
\frac{\partial}{\partial z}   \hat \delta_s  (z )   =  -\frac{ g_0}{\phi }  
 \hat \delta_i^\dagger(z)  e^{-i \D(\w_s) z }
\\[0.9em]
\frac{\partial}{\partial z}   \hat \delta_i  (z )   =  -\frac{ g_0}{\phi }  
 \hat \delta_s^\dagger ( z)  e^{-i \D (\w_s) z }
\end{cases}
\end{align}
Notice that  the transformation \eqref{trphi} is canonical, i.e. such that  $\hat \sigma_j$ and $\hat \delta_j$ are independent photon annihilation operators, and that the two linear systems  \eqref{sigmaprop} and \eqref{deltaprop} are the standard  equations describing parametric amplification in a pair of conjugate signal-idler modes. Thus,  each of them  generates a pair of independent twin beams,  where 
modes $\hat \sigma_s, \hat \sigma_i $  are characterized by an  enhanced  gain/squeeze  parameter: 
\beq
g_0 \to g_0 \times \phi = g_0 \times 1.618..., 
\eeq
 whereas   the gain/squeeze parameter  is reduced for modes $ \hat \delta_s,  \hat \delta_i$:  
\beq
g_0 \to - \frac{g_0 }{\phi}= - g_0 \times 0.618...
\eeq
Their  explicit solutions can be written  within  the input-output formalism (Appendix \ref{AppA}) as: 
\bsub
\label{inout2}
\begin{align}
& \sigmas^{\mathrm out}  = U_{\phi}  (\w) \, \sigmas^{\mathrm in}  + V_{\phi} (\w) \, \sigmai^{\dagger \, in}
\label{inoutsigma}\\
& \deltas^{\mathrm out}  = U_{-\frac{1}{\phi}}  (\w) \, \deltas^{\mathrm in}  + V_{-\frac{1}{\phi}}   (\w) \, \delta_i^{\dagger \, in}
\label{inoutdelta}
\end{align}
and analog transformations for the output idlers,  obtained by exchanging  the indexes $ s \leftrightarrow i$. 
These solutions have to be considered together with the inverse transformation 
\begin{align}
\begin{pmatrix} 
\hat b_j \\[0.8em]
\hat c_j
\end{pmatrix} 
=  
%\frac{ 1}{\sqrt{1+\phi^2}} 
\begin{pmatrix}
\frac{ 1}{\sqrt{1+\phi^2}} \; &  \frac{ \phi}{\sqrt{1+\phi^2}}  \\[0.8em]
\frac{ -\phi}{\sqrt{1+\phi^2}} \; &  \frac{ 1}{\sqrt{1+\phi^2}}
\end{pmatrix} 
\begin{pmatrix} 
\hat \delta_j \\[0.8em]
\hat \sigma_j
\end{pmatrix} 
\qquad (j=s,i) 
\label{trphi2}
\end{align} 
\esub
 We notice that   the original modes $\hat b_j $, $\hat c_j$   can be obtained from modes $\hat \sigma_j$ and $\hat \delta_j$  (and viceversa) via the action of a non balanced beam splitter with reflection and transmission coefficients 
$
t= \frac{1}{\sqrt{1+ \phi^2}} $, $r = \frac{\phi}{\sqrt{1+ \phi^2}}$, as 
$
 \hat b_j  =  t \delta_j  +  r \hat \sigma _j   $, 
$\hat c_j =  - r \hat  \delta_j  + t\hat \sigma_j  $. 
Therefore, in the resonant case, the 4-mode entanglement can be considered equivalent to (see Fig. \ref{figgolden}):
\begin{itemize}
\item Two independent parametric processes, with different gain parameters $g_0 \phi$ and $-{g_0} /{\phi} $, generating  two  independent pairs of entangled twin- beams in modes $\sigma_j$ and $\delta_j$;  
\item Followed by a non-balanced beam splitter that performs a "golden partition" of the two twins into the original modes 
\end{itemize}
We thus see that a resonant pump provides the maximal coherence between the two nonlinear processes that coexist in the HexNPC .
%matched by  vectors $\Gone$ and $\Gtwo$ of the nonlinear lattice.  
Indeed, away from this resonance, the 3-mode entanglement generated by the HexNPC is equivalent to a single nonlinear process (as depicted in Fig.\ref{fig_solution}) generating bipartite entanglement, followed by a beam-splitter acting on one of the two parties. Conversely, when  the pump resonates with the lattice, the output of the  device  is equivalent  to {\em two} independent nonlinear processes generating two  bipartite entangled  states,  followed by a linear device that mixes them. \\
%%%%%%%%%
%%%%%
This maximal coherence is also reflected into an enhancement of the intensity of hot-spots at resonance, experimentally  observed  in \cite{Gallo2018}. The mean photon numbers can be calculated from the solutions \eqref{inout2} as 
\begin{align}
\langle \hat b_j^{\dagger \, out} \hat b_j^{ out}  \rangle & =   \frac{\phi^2}{1+\phi^2} \langle \hat \sigma_j^{\dagger \, out} \hat \sigma_j^{ out}  \rangle + \frac{1}{1+\phi^2} \langle \hat \delta_j^{\dagger \, out} \hat \delta_j^{ out}  \rangle  \nn \\
&=\delta (0) \, \left[ \frac{\phi^2}{1+\phi^2}  \left| V_\phi (\w_s) \right|^2 +     \frac{1}{1+\phi^2} \left| V_{-\frac{1}{\phi}} (\w_s) \right|^2 \right] \label{intb} \\
& \to \delta (0) \frac{\phi^2}{1+\phi^2}  \sinh^2 \left( g_0 \phi l_c\right)  \qquad \qquad \text{for} \quad  g_0 l_c \gg 1
\end{align}
\begin{align}
\langle \hat c_j^{\dagger \, out} \hat c_j^{ out}  \rangle & =   \frac{1}{1+\phi^2}   \langle \hat \sigma_j^{\dagger \, out} \hat \sigma_j^{ out}  \rangle + \frac{\phi^2}{1+\phi^2} \langle \hat \delta_j^{\dagger \, out} \hat \delta_j^{ out}  \rangle  \nn \\
&=\delta (0) \, \left[ \frac{1}{1+\phi^2}  \left| V_\phi (\w_s) \right|^2 +     \frac{\phi^2}{1+\phi^2} \left| V_{-\frac{1}{\phi}} (\w_s) \right|^2 \right] \label{intc} \\
& \to \delta (0) \frac{1}{1+\phi^2}  \sinh^2 \left( g_0 \phi l_c \right)  \qquad \qquad \text{for} \quad  g_0 l_c \gg 1
\end{align}
where the last lines hold for phase-matched modes $\D=0$ and  in the regime of  stimulated PDC,   i.e  for $g_o l_c \gg 1$, where  the  mode intensities  grow exponentially with $g_0 z$. In such conditions,   the contribution  of modes  $\delta $ with smaller gain becomes negligible, and,  as observed in \cite{Gallo2018} the local enhancement of  intensity in the hot-spots is ruled by the Golden ratio: $I_{4-modes} \simeq e^{2g_0 \phi l_c} = ( I_{\rm background})^\phi$.  The Golden Ratio also rules the  ratio between the amplitudes  of the shared and unshared modes because 
$  \langle \hat b_j^{\dagger \, out} \hat b_j^{ out}  \rangle/ \langle \hat c_j^{\dagger \, out} \hat c_j^{ out}  \rangle \to \phi^2$.  \\
Conversely, in the spontaneous  PDC limit,  by using the asymptotic behaviour  of the functions $\Vg$ provided in  Appendix \ref{AppA},  the expressions \eqref{intb}, \eqref{intc} reduce to Eqs\eqref{int_low}, where the Golden ratio does not appear and the ratio of the intensities of  the shared and unshared mode is just 2. 
In this connection, it is also interesting to  consider the quantum  state generated by the HexNPC  in the same limit This can be written in general as the state generated from the vacuum by the action of the two-mode squeeze operators acting on modes $\sigma_j$ and $\delta_j$ plus the action of the beam-splitter. As this is a rather cumbersome formula,  we limit to its expression in  the spontaneous regime ($ g_0 l_c\ll 1$): 
\begin{align}
| \psi_{out}\rangle& = \hat B_{\delta \sigma} \hat R_{\delta_s \delta_i} \left(\xi_{-1/\phi} \right) 
\hat R_{\sigma_s \sigma_i} \left(\xi_{\phi} \right)| \psi _{in}\rangle \nn \\
 &  \stackrel{ g_0 l_c \ll 1} {\longrightarrow} \, 
 |0\rangle + g_0 l_c\, \sinc \left(\frac{D(\w_s) l_c}{2} \right)\, e^{-i \frac{D (\w) l_c}{2} }
\left[ \hat b^\dagger_s \hat c^\dagger_i 
+ \hat b^\dagger_s \hat b^\dagger_i 
+ \hat c^\dagger_s \hat b^\dagger_i \right] \; 
|0\rangle 
\end{align} 
In this equation the contributions of the three microscopic processes allowed by energy-momentum conservation depicted in Fig.\ref{figinteractions} are evident. Since they take place with the same probability, in the spontaneous regime  
where the elementary down-conversion processes occur independently, the ratio of the two intensities is 2, and the Golden Ratio does not appear. This number appears only asymptotically, as the chain of stimulated processes become longer and longer. 
\par
One can recognize here an analogy with the Fibonacci sequence, defined by the recurrence relation $F_{n+1}= F_{n} + F_{n-1}$, with initial values $F_1=F_2=1$. As well known \cite{Dunlap2017},  the ratio between two consecutive numbers $\frac{F_{n+1}}{ F_{n}} $ goes asymptotically to the Golden Ratio $\phi$  for $n\to \infty$.  Indeed, by considering the $D=0$ case,   introducing proper hermitian quadrature operators,  and the normalized distance $\bar z= g_0 z$,  the dynamical evolution of the { amplitudes} of modes $b$ and $c$ in Eqs. \eqref{resonance_prop} 
%belongs to a general  class of popolation growth model 
can be reformulated as  the simpler model
\footnote{Precisely, these equations describe the evolution of the most amplified field quadratures in the resonant case 
$\hat B = \frac{1} {\sqrt 2} \left (\hat b_s + \hat b_i + \hat b_s^\dagger + \hat b_i^\dagger\right) $, $\hat C= \frac{1} {\sqrt 2} \left (\hat c_s + \hat c_i + \hat c_s^\dagger + \hat c_i^\dagger\right) $.} 
\beq
\label{BC}
\begin{array} {rl} 
\frac{d B }{d\bar z}  &= B+C \\[0.3em]
\frac{d C}{d\bar z}   &= B 
\end{array} 
\eeq
whose eigenvelues $\phi$ and $-\frac{1}{\phi}$ are the solution of  the quadratic characteristic equations $\lambda^2=\lambda + 1$.  It is possible to write a  discrete version of the continuous evolution in Eq.\eqref{BC},  introducing 
 $B_n .= B(n \Delta z) , C_n= C (n \Delta z)$, where  $\Delta z = z/n$ is a discrete step.   It can be then demonstrated that  at the n-th layer the amplitudes obey the Fibonacci-like recursive relations
$B_{n+1} = (2+ \Delta z) B_n + (\Delta z^2 -1- \Delta z) B_{n-1}$. 
% which admit geometrical solutions of the form  $B_n = \lambda_1^n k_1 + \lambda_2^n k_2 $, where the  eigenvalues are $\lambda_1 = 1 + \Delta z  \phi$ and  $\lambda_2 = 1 -\Delta z / \phi$.
A similarity then holds with  the famous description of the evolution of the population of rabbits introduced  in the 13th century by L. Fibonacci \cite{Koshy2011ch2}
\beq
\label{FN} 
\begin{array} {rl} 
F_ {n+1}  &= F_{n} + N_{n} \\
N_{n+1}  &=  F_{n}
\end{array} 
\eeq
where $F_n$ and $N_n$ are the number of adult and newly born rabbits at month $n$, respectively.  
With initial conditions $N_1=0, F_1=1$, the solution is  the Fibonacci sequence $F_n= \frac{\phi^n -
 \left(-{1}/ {\phi} \right)^n }{\sqrt 5}$. Clearly the analogy is limited, also  because  the quadratures  in Eqs.\eqref{BC} are  not integer, since they are initiated by vacuum  fluctuations, but the two models share the same eigenvalues (the Golden Ratio and its inverse),  exhibiting  an exponential and a  geometrical growth rate, in the case of  Eq.\eqref{BC} and Eq.\eqref{FN}),  respectively. 
Interestingly, the  asymptotic  behaviour of the ratio between the two  variables is also preserved: 
$\lim_{\bar z \to \infty} \frac{B(\bar z) }{C(\bar z}  = \lim_{ n \to \infty} \frac{F_n}{N_n}=  \phi$.
%%%%%%%%%
\begin{figure}[ht]
\includegraphics[scale=0.6]{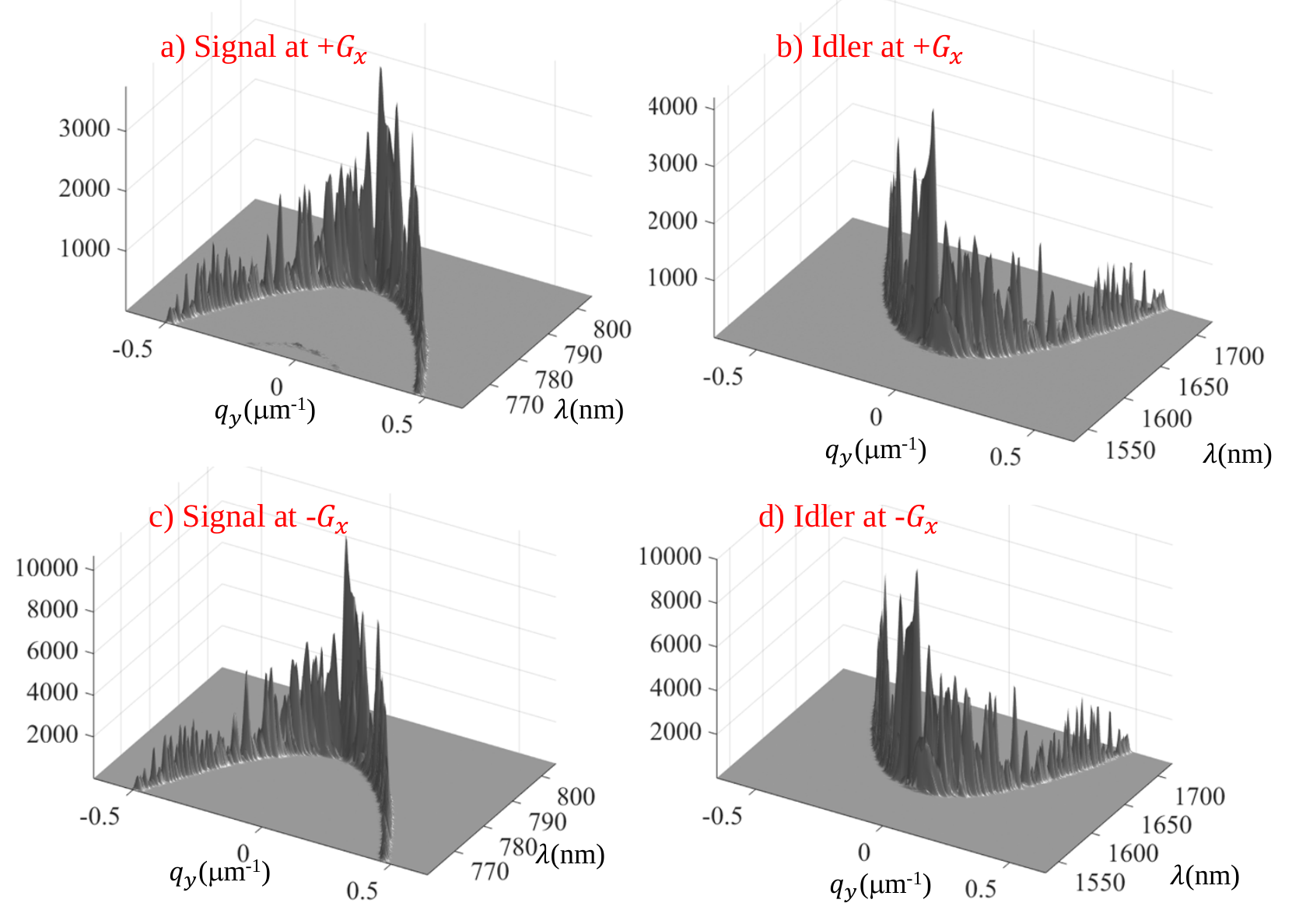}
\caption{Spectra of the hot-spots at superresonance $(q_p=-G_x)$ as  the $q_y$ coordinate is varied. Numerical simulations of 
the photon-number distributions in the $(q_y, \lambda)$ plane at $q_x= +G_x$ (a,b) and $q_x=-G_x$ (b,c).  Other parameters as in Fig.\ref{figHS}. 
}
\label{figspectraqy}
\end{figure}
%%%%%
\par
As already remarked for the 3-mode case, also the Golden Ratio entanglement is widely tunable. Indeed the 4-mode entangled state  is  generated for a continuous range of shared signal modes $\w_s$, spanning a broad band of frequencies as the $q_y$ coordinate is scanned along the hot-spot lines.  This is evidenced by Fig. \ref{figspectraqy} which show numerical simulations of  the spectra of the hot-spots as the $q_y$ coordinate is varied. 
\section{Conclusions}
In this work, we studied the quantum state of twin photons and twin beams generated by parametric down-conversion in a hexagonally poled photonic crystal. The latter is characterized by the coexistence of two nonlinear processes, and represents an interesting monolithic source of path entangled photonic states. In particular we presented a novel point of view of the tripartite state that can be produced and we described the peculiar 4-mode entangled state generated by proper angular tuning of the input pump, in a regime we refer to as superresonance. 
\par
Away from superresonance, our description generalizes and partially confirms the analysis performed by other authors [5] in the low gain regime. Indeed our analysis of the 3 mode entanglement, not limited to the two-photon state, shows that it is necessary to perform a conditional measurement on the shared mode to produce quantum entanglement in the other two coupled modes. In the absence of such a conditional measurement, the state of the coupled modes is a non-entangled state, exhibiting the classical correlation/coherence properties of splitted thermal beams. Therefore in spite of the interesting compactness of the device, we can see in this case a somehow trivial analogy with the output state from a beam-splitter. \\
In contrast, we have shown that the angular degrees of freedom  of the pump can be used for a further engineering of the output state. In particular at superresonance, the 4-mode state thereby generated is definitely non-trivial, as the spatial resonance of the pump with the lattice establishes a strong coherence between the two nonlinear processes that cohexist in the hexagonal photonic crystal. The output of the device can be seen as two bipartite entangled states followed by a linear device that performs a golden partition. The curious appearance of the golden ratio in this physical system can be explained by recognizing the existence of an analogy between the dynamical evolution of the mode amplitudes and Fibonacci sequence.
\par
Finally,  the original aspect of our theoretical model has allowed to highlight the spectral and spatial tunability of the HexNPC, as the entanglement described is generated over a large frequency bandwidth and over a continuous range of Fourier modes, rendering this compact monolithic device appealing for different integrated quantum optics and optical parametric generation experiments, that may require a versatility in the generation or detection schemes. 
%%%%%%%%%
\appendix
\section{Solutions of coupled propagation equations} 
\label{AppA}
Let us consider the generic 2-mode coupled parametric equations 
\begin{align}
&\frac{\partial }{\partial z} \hat a_s (\w,z) = \gamma g_0 \hat a_i^\dagger (-w,z) e^{-i \D (\w) z} \\
&\frac{\partial  }{\partial z} \hat a_i (-\w,z) = \gamma g_0 \hat a_s^\dagger (w,z) e^{-i \D (\w) z} 
\end{align}
where $\gamma $ is a gain enhancement factor, arising from a coherent superposition of the two nonlinear processes characteristic of the hexagonal nonlinear pattern: $\gamma=1$ for the background fluorescence (no coherence between the two processes) , $\gamma= \sqrt 2$ for the 3-mode coupling ;  $\gamma= \phi$ or $\gamma= -1/\phi$ for the 4-mode coupling at resonance. 
In terms of input field operators, their solution at the crystal exit face $z=l_c$ can be written as 
\begin{align}
\hat a_s (\w,l_c) = \Ug (\w)  \hat a_s (\w,0) + \Vg (\w) \hat a_i^\dagger (-\w,0)  \\
\hat a_i (-\w,l_c) = \Ug (\w)  \hat a_i (-\w,0) + \Vg (\w) \hat a_s^\dagger (\w,0) 
\end{align}
Alternativa 
\begin{align}
\hat a_s (\w, z) = \Ug (\w,z)  \hat a_s (\w,0) + \Vg (\w,z) \hat a_i^\dagger (-\w,0)  \\
\hat a_i (-\w, z) = \Ug (\w,z)  \hat a_i (-\w,0) + \Vg (\w,z) \hat a_s^\dagger (\w,0) 
\end{align}
where 
\begin{align}
\Ug (\w) &=\left[  \cosh \left[ \Gamma (\w)l_c \right]  + \frac{i \D(\w) } {2\Gamma(\w)} 
\sinh   \left[ \Gamma (\w) l_c \right]     \right]  e^{-i  \frac{\D (\w)l_c}{2} }  \nn \\
\Vg (\w) &=\gamma g_0 \frac{\sinh   \left[ \Gamma (\w) l_c\right] }{\Gamma (\w)}  e^{-i  \frac{\D (\w)l_c}{2} } 
\label{UV}\\
\Gamma (\w) &= \sqrt{ |\gamma g_0|^2 -\frac{\D^2(\w)}{4} } 
\nn
\end{align}
In the low-gain limit  where $\gamma g_0 l_c \to 0$, with $\D l_c$ finite (spontaneous PDC limit), the coefficients take the familiar form 
$\Ug (\w) \to 1$ and 
$\Vg (\w)  \to \gamma g_0 l_c \sinc  {\left[   \frac{\D (\w)l_c}{2} \right] }    e^{-i  \frac{\D (\w)l_c}{2} } $. \\
If instead of evolving the field along the sample, one chooses to work in the picture where the state evolves, then the output state is given by the action of the two-mode squeeze operator on the input state: 
\begin{align}
|\psi_{out}\rangle &= \hat R (\xi) \, |\psi_{in}\rangle  \nn \\
\hat R (\xi) &= e^{\xi (\w) \hat a_s ^\dagger  (\w)\hat a_i ^\dagger (-w)-\xi (\w)^*\hat a_s (\w)  \hat a_i (-w)}
\end{align}
where the squeeze parameter $\xi= r e^{2 i \theta} $ is determined by  the input-output coefficients as 
$\tanh (r) = \frac{|\Vg|}{ |\Ug|}$, $\theta = \frac{1}{2} {\rm arg} \left[ \Ug \Vg \right] $. Notice that for phase matched modes for which $\D(w) =0$,  it simply reduces to $\xi= \gamma g_0 l_c$.

---------------------------------\\

\bibliography{biblio_HEXNPC2018}
\bibliographystyle{apsrev4-1}
\end{document}